\documentclass[twocolumn,amsmath,amssymb]{revtex4}
\input{epsf}

\usepackage{graphicx}
\usepackage{array}
\usepackage{bigstrut}
\usepackage{booktabs}
\usepackage{longtable}
\usepackage{rotating,booktabs}
\usepackage{booktabs,threeparttable}
\usepackage{bm}
\usepackage{lipsum}
\usepackage{braket}
\usepackage{amsmath}
\usepackage{multirow}
\usepackage{bm}
\usepackage{threeparttable}
\usepackage{color}
\usepackage{placeins}
\usepackage{pdfpages}

\begin{document}

\title{Simultaneous Determination of Multiple Nuclear Parameters of $^{229}$Th Using Highly Charged Ions}

\author{ Hong-Yuan Zheng$^{1,2,*}$, Yan-Ling Xu$^{1,2,*}$, Xi-Chen Yu$^{1,2}$,  Yong-Hui Zhang$^{1, \dag }$, Zong-Chao Yan$^{3,1}$, Li-Yan Tang$^{1,\S}$, and Xiaojun Liu$^{1}$~\footnotetext{ *\,These authors contributed equally to this work.\\ \dag\,yhzhang@apm.ac.cn \\  $\S$\,lytang@apm.ac.cn}}
	
	\affiliation {$^1$ State Key Laboratory of Magnetic Resonance and Atomic and Molecular Physics, Wuhan Institute of Physics and Mathematics, Innovation Academy for Precision Measurement Science and Technology, Chinese Academy of Sciences, Wuhan 430071, People's Republic of China}
	
	\affiliation {$^2$ University of Chinese Academy of Sciences, Beijing 100049, People's Republic of China}
	
	\affiliation {$^3$ Department of Physics, University of New Brunswick, Fredericton, New Brunswick, E3B 5A3, Canada}
	
	\date{\today}
	
	\begin{abstract}		
 Development of a $^{229}$Th nuclear optical clock requires precise and model-insensitive nuclear-structure parameters, which presently suffer from limited accuracy and poor consistency. We propose a joint spectroscopy scheme using two highly charged $^{229}$Th ions with $J=1/2$ electronic ground states, where the lowest electronic excitation energy of each ion far exceeds the nuclear transition energy.
 This configuration effectively forms a three-level system comprising the electronic ground state and the nuclear ground ($g$) and isomeric ($m$) states, resulting in strongly enhanced nuclear hyperfine mixing. 
Within this framework, a unified analysis of precision measurements on both ions enables the simultaneous determination of five key nuclear parameters without relying on external nuclear inputs: the magnetic dipole moments $\mu_g$ and $\mu_m$, the bare-nucleus transition energy $\omega_n$, the charge-radius difference $\delta\langle r^{2}\rangle_{gm}$, and the $M1$ transition matrix element $T_{M1}$. With this approach, the uncertainties in $\omega_n$ and $\delta\langle r^{2}\rangle_{gm}$ are estimated to be reduced by factors of 3 and 2, respectively, relative to their current uncertainties. 
This work could provide a useful benchmark for nuclear theory and serve as a foundation for future development of a $^{229}$Th-based nuclear optical clock. 
\end{abstract}
	
	\maketitle
	
\textit{Introduction}—The $^{229}$Th nucleus hosts a uniquely low-lying isomeric state, making it the only system currently amenable to laser manipulation for a nuclear optical clock of exceptional precision~\cite{peik03a,zhang24,maheshwari24,vonderwense20}. Such a clock promises major advances in fundamental physics and metrology, ranging from searches for dark matter and variations of fundamental constants to a potential redefinition of the second~\cite{Thirolf24a,peik2021,caputo25,fadeev20,zaheer25}. Its realization, however, requires precise knowledge of several key nuclear parameters, including the magnetic dipole moments ($\mu_g$, $\mu_m$), the bare-nucleus transition energy $\omega_n$, the charge-radius difference $\delta\langle r^{2}\rangle_{gm}$, and the magnetic-dipole reduced transition probability $B(M1)$.

Current knowledge of these parameters remains insufficient. The value of $\omega_n$ inferred from $^{229}$Th$^{4+}$-doped crystals carries an uncertainty of several THz~\cite{perera25a}, while $\delta\langle r^{2}\rangle_{gm}$ is known with only about 12\% precision~\cite{safronova18a}. The situation is most acute for $B(M1)$, which governs the radiative lifetime of the isomer and is determined by the nuclear $M1$ matrix element $T_{M1}$~\cite{porsev10a,minkov17,bilous18b}. Experimentally inferred values of $B(M1)$ span nearly an order of magnitude~\cite{porsev10a,tiedau24,shabaev22}. Neither \textit{ab initio} nuclear theory nor direct spectroscopy has yet achieved the required accuracy~\cite{minkov17,kozlov24}.

This impasse has motivated indirect approaches based on precision atomic spectroscopy~\cite{beloy14,porsev10}, notably exploiting nuclear hyperfine mixing (NHM) in highly charged ions (HCIs) such as $^{229}\mathrm{Th}^{87+}$ to extract nuclear parameters~\cite{shabaev22}. However, the accuracy of these methods is inherently limited by their reliance on additional nuclear inputs that are themselves imperfectly known. Another obstacle is the Bohr--Weisskopf (BW) effect~\cite{bohr50,shabaev99hfs}, which accounts for the finite spatial distribution of nuclear magnetization. For HCIs, the associated uncertainty can reach $\sim$35\%, constituting the dominant source of error in calculated hyperfine matrix elements~\cite{tkalya16,shabaev22}. 

In this Letter, we introduce an overdetermined spectroscopic scheme that enables the simultaneous determination of five key nuclear parameters without relying on external nuclear inputs. Our method exploits the strongly enhanced NHM effect in selected HCIs ($^{229}\mathrm{Th}^{87+}$ and $^{229}\mathrm{Th}^{79+}$), where the first electronic excitation energy far exceeds the nuclear isomer energy. In this regime, the second-order hyperfine shift is dominated by nuclear intermediate states, leading to a strong enhancement of NHM. From six measurable hyperfine transitions, we derive a closed set of equations that can be analytically inverted to obtain unique solutions for $\mu_g$, $\mu_m$, $\omega_n$, $\delta\langle r^{2}\rangle_{gm}$, and $T_{M1}$, provided the electronic quantities are determined with sufficient accuracy. This approach provides a self-contained path to extracting the nuclear parameters necessary for the $^{229}\mathrm{Th}$ nuclear clock, thereby helping to establish a robust foundation for its future realization.

\begin{figure}[htbp] 
	\includegraphics[width=0.455\textwidth]{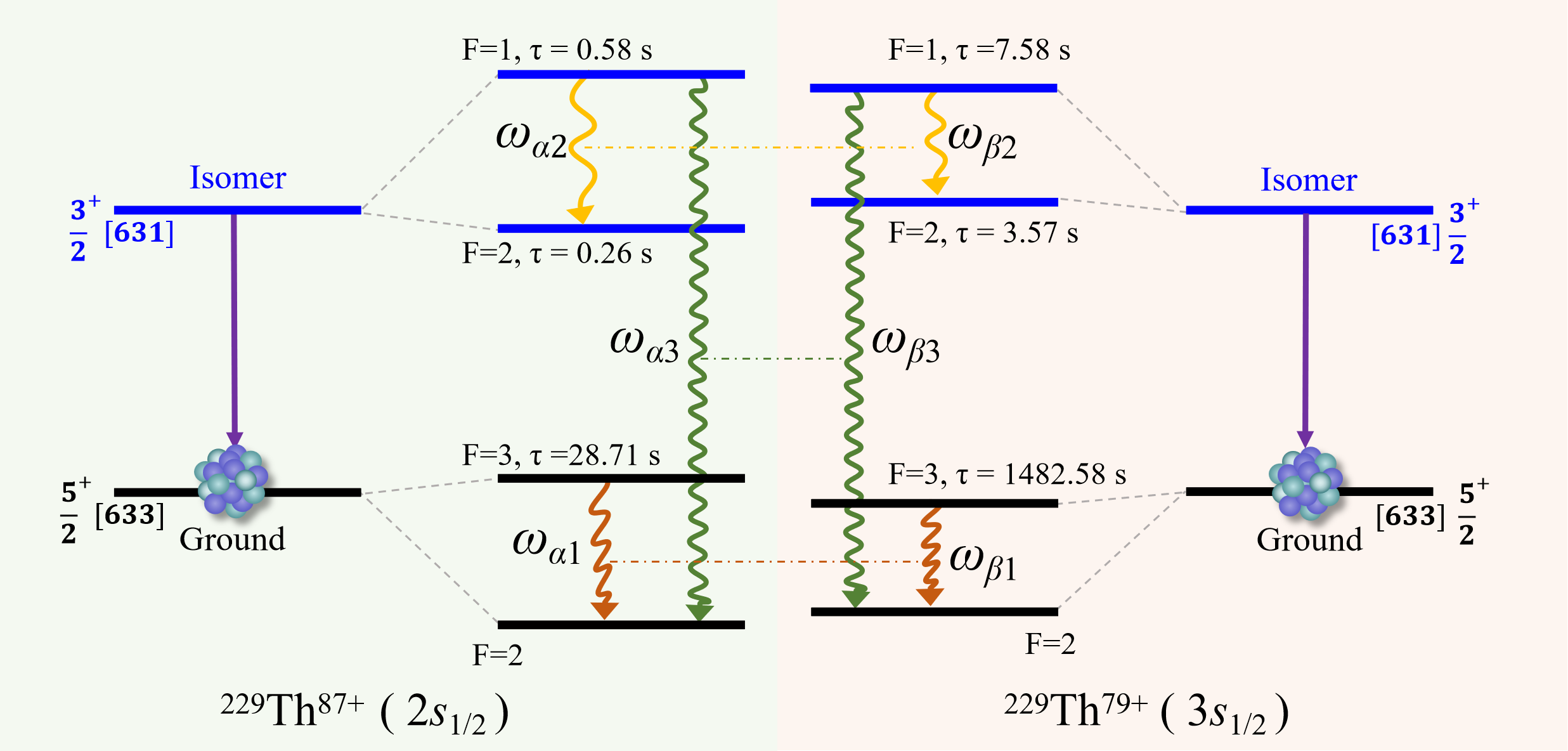}
	\caption{Combined spectroscopic scheme based on NHM transitions in $^{229}\mathrm{Th}^{87+}$ and $^{229}\mathrm{Th}^{79+}$ ions for extracting nuclear parameters.}
	\label{fig1} 
\end{figure}

\textit{Combined spectroscopic scheme for extracting nuclear parameters}—Our scheme exploits a distinctive regime inherent in HCIs such as $^{229}\mathrm{Th}^{87+}$ and $^{229}\mathrm{Th}^{79+}$, each with a $J=1/2$ valence electron. In these systems, the first electronic excitation energy is much greater than the nuclear isomer energy $\omega_n$, effectively restricting the hyperfine interaction to three levels: the electronic ground state and the nuclear ground ($g$) and isomeric ($m$) states. In this configuration, the second-order hyperfine energy shift is dominated by the nuclear isomeric state, resulting in an enhancement of the NHM effect by about four orders of magnitude. Fig.~\ref{fig1} illustrates the level structure and the relevant transition energies. 


Specifically, the hyperfine splitting of the electronic $2s_{1/2}$ state in $^{229}\mathrm{Th}^{87+}$ and the $3s_{1/2}$ state in $^{229}\mathrm{Th}^{79+}$ produces six independent transitions associated with the nuclear spins $I_g = 5/2$ and $I_m = 3/2$.
Their frequencies, including first-order magnetic and second-order NHM contributions (see Secs.~S1–S3 of the Supplemental Material), are:
\begin{align}
	\omega_{\alpha1} &=
	\frac{2\sqrt{6}}{5}\,\mu_gT_{g, 2s}
	+\frac{T_{M1}^2 T_{g, 2s}^2}{15\,\omega_n},
	\label{wa1} \\[2pt]
	\omega_{\alpha2} &=
	-\frac{4\sqrt{6}}{9}\,\mu_mT_{m, 2s}
	-\frac{T_{M1}^2 T_{m, 2s}^2}{15\,\omega_n},
	\label{wa2} \\[2pt]
	\omega_{\alpha3} &=
	\omega_n + F^{\mathrm{ved}}_{2s}\,\delta\langle r^{2}\rangle_{gm}
	-\frac{5\sqrt{6}}{18}\,\mu_mT_{m, 2s}
	+\frac{7\sqrt{6}}{30}\,\mu_gT_{g, 2s} \nonumber\\
	&\quad +\frac{T_{M1}^2T_{g, 2s}^2 }{15\,\omega_n},
	\label{wa3} \\[2pt]
	\omega_{\beta1} &=
	\frac{2\sqrt{6}}{5}\,\mu_gT_{g, 3s}
	+\frac{T_{M1}^2 T_{g, 3s}^2}{15\,\omega_n},
	\label{wb1} \\[2pt]
	\omega_{\beta2} &=
	-\frac{4\sqrt{6}}{9}\,\mu_mT_{m, 3s}
	-\frac{T_{M1}^2 T_{m, 3s}^2}{15\,\omega_n},
	\label{wb2} \\[2pt]
	\omega_{\beta3} &=
	\omega_n + F^{\mathrm{ved}}_{3s}\,\delta\langle r^{2}\rangle_{gm}
	-\frac{5\sqrt{6}}{18}\,\mu_mT_{m, 3s}
	+\frac{7\sqrt{6}}{30}\,\mu_gT_{g, 3s} \nonumber\\
	&\quad +\frac{T_{M1}^2T_{g, 3s}^2 }{15\,\omega_n}.
	\label{wb3}
\end{align}

These six equations involve five unknown nuclear parameters ($\mu_g$, $\mu_m$, $\omega_n$, $\delta\langle r^2\rangle_{gm}$, and $T_{M1}$), together with several calculable atomic properties: the nuclear-state-dependent electronic reduced matrix elements $T_{g(m),2s(3s)}$ and the effective field-shift factors $F_{2s(3s)}^{\rm ved}$. With six equations for five unknowns, the system is overdetermined, allowing internal consistency checks. To exploit this overdeterminacy, we derive a constraint equation by combining Eqs.~(\ref{wa1}), (\ref{wa2}), (\ref{wb1}), and (\ref{wb2}), in which the nuclear unknowns cancel identically:
\begin{eqnarray}
\frac{\omega_{\alpha 2}}{T_{m,2s}} - \frac{\omega_{\beta 2}}{T_{m,3s}} 
	= -\,\frac{T_{m,2s} - T_{m,3s}}{T_{g,2s} - T_{g,3s}} 
	\left( \frac{\omega_{\alpha 1}}{T_{g,2s}} - \frac{\omega_{\beta 1}}{T_{g,3s}} \right).&&
	\label{eq_consistency}
\end{eqnarray}

Under this constraint, the five independent nuclear parameters can be expressed analytically as:
\begin{align}
	\mu_g &=
	\frac{5\sqrt{6}\left(T_{g,2s}^{2}\omega_{\beta1}-T_{g,3s}^{2}\omega_{\alpha1}\right)}
	{12\,T_{g,2s}T_{g,3s}\left(T_{g,2s}-T_{g,3s}\right)},
	\label{ex_mug}
	\\[2pt]
	\mu_m &=
	\frac{3\sqrt{6}\left(T_{m,2s}^{2}\omega_{\beta2}-T_{m,3s}^{2}\omega_{\alpha2}\right)}
	{8\,T_{m,2s}T_{m,3s}\left(T_{m,3s}-T_{m,2s}\right)},
	\label{ex_mum}
	\\[2pt]
	\omega_n &=
	\frac{F_{2s}^\mathrm{ved}\left(\omega_{\beta3}-R_2\right)-F_{3s}^\mathrm{ved}\left(\omega_{\alpha3}-R_1\right)}
	{F_{2s}^\mathrm{ved}-F_{3s}^\mathrm{ved}},
	\label{ex_wn}
	\\[2pt]
	\delta\langle r^{2}\rangle_{gm} &=
	\frac{\omega_{\alpha3}-\omega_{\beta3}+R_2-R_1}{F_{2s}^\mathrm{ved}-F_{3s}^\mathrm{ved}},
	\label{ex_r2}
	\\[2pt]
	T_{M1} &=
	\sqrt{\frac{15\left(\omega_{\alpha1}T_{g,3s}-\omega_{\beta1}T_{g,2s}\right)}
		{T_{g,2s}T_{g,3s}\left(T_{g,2s}-T_{g,3s}\right)}}
	\nonumber\\
	&\times
	\sqrt{\frac{F_{2s}^\mathrm{ved}\left(\omega_{\beta3}-R_2\right)-F_{3s}^\mathrm{ved}\left(\omega_{\alpha3}-R_1\right)}
		{F_{2s}^\mathrm{ved}-F_{3s}^\mathrm{ved}}},
	\label{ex_x}
\end{align}
where $R_1$ and $R_2$ are given by
\begin{align}
	R_1=&\frac{\left( T_{g,3s}\omega _{\alpha 1}-T_{g,2s}\omega _{\beta1} \right) \left( 39T_{m,2s}^{2}-14T_{g,2s}^{2} \right)}{24T_{g,2s}T_{g,3s}\left( T_{g,2s}-T_{g,3s} \right)}\nonumber\\
	&+\frac{14\omega _{\alpha 1}+15\omega _{\alpha 2}}{24},
	\\[2pt]
	R_2=&\frac{\left( T_{g,3s}\omega _{\alpha 1}-T_{g,2s}\omega _{\beta1} \right) \left( 39T_{m,3s}^{2}-14T_{g,3s}^{2} \right)}{24T_{g,2s}T_{g,3s}\left( T_{g,2s}-T_{g,3s} \right)}\nonumber\\
	&+\frac{14\omega _{\beta1}+15\omega _{\beta2}}{24}.
\end{align} 
%
%
These solutions are fully determined by the measured transition frequencies and calculable atomic properties, without external nuclear inputs. In addition, Eq.~(\ref{eq_consistency}) provides a stringent internal consistency check on the combined experimental and theoretical inputs. Since this relation is sensitive to the BW effect and higher-order QED corrections, its fulfillment constitutes a nontrivial validation of the framework.

\textit{Evaluation of uncertainty in nuclear parameters}—{ Hyperfine transition measurements in these HCIs are experimentally feasible (see Sec.~S6 of the Supplemental Material (SM)), but no such data for \(^{229}\mathrm{Th}^{87+}\) and \(^{229}\mathrm{Th}^{79+}\) are currently available. We therefore calculate the six transition frequencies together with the corresponding atomic structure properties to assess the reliability of our scheme; the results are listed in Table~\ref{inputs}. The \(\beta_1\) and \(\beta_2\) splittings lie in the 7--8~THz range, where direct THz fluorescence detection is challenging; however, they can be accessed through VUV difference-frequency measurements or frequency-comb Raman spectroscopy, as discussed in the SM. Based on the precision demonstrated for Bi\(^{80+}\) ions~\cite{ullmann17}, we conservatively assign a 5~GHz uncertainty to these frequencies, which is far larger than the kHz-level precision already achieved in doped \(^{229}\mathrm{Th}^{4+}\) crystals~\cite{zhang24}. Other atomic properties, such as the \(M1\) radiative lifetimes of the relevant hyperfine states (including NHM-induced decay channels), are listed in Sec.~S4 of the SM. The hyperfine-state lifetimes are \(0.26\)--\(28.71\)~s for \(^{229}\mathrm{Th}^{87+}\) and \(3.57\)--\(1482.58\)~s for \(^{229}\mathrm{Th}^{79+}\), corresponding to natural linewidths below \(1\)~Hz and therefore not limiting the GHz-level spectroscopic precision assumed here.}

The dominant theoretical uncertainty arises from the electronic matrix elements $T_{g(m),2s(3s)}$, which are affected by the BW effect. To mitigate this source of uncertainty, we adopt the ratio-based approach proposed in Ref.~\cite{shabaev22}, which exploits the strong correlation of 
BW effects across different charge states and $ns$ configurations. Specifically, the BW-corrected matrix elements are normalized to a common hydrogen-like BW reference, and the resulting ratios are used to parametrize these matrix elements. This procedure significantly reduces the BW-related uncertainty in Eqs.~(\ref{ex_mug})--(\ref{ex_x}). 

The BW-related inputs, namely the ratio factors \(R_{g(m),2s(3s)}\) and the hydrogen-like BW corrections \(\epsilon_{g(m),1s}\), are taken from Refs.~\cite{shabaev22,shabaev01,roberts22}. {We adopt the approximation $R_{g(m),3s} = R_{g(m),2s}$, and tests of this approximation confirm its validity for the uncertainty evaluation in our scheme (see Sec. S5 of the SM).} The uncorrected electronic matrix elements \(T^{(0)}_{2s(3s)}\) (i.e., without BW corrections) and the field-shift factors \(F^{\mathrm{ved}}_{2s(3s)}\) are obtained from our multiconfiguration Dirac–Hartree–Fock (MCDHF) calculations performed with the GRASP2018 and RIS4 packages~\cite{ekman19a,Fischer19a}. All input parameters are summarized in Table~\ref{inputs}.

\begin{table*}[!htbp] 
	\tabcolsep=0.12cm
	\renewcommand\arraystretch{1.12}  
\caption{\label{inputs}
	Input values and predicted nuclear parameters. 
	The six hyperfine transition frequencies are obtained from atomic-structure 
	calculations. Atomic-structure quantities 
	and BW corrections~\cite{shabaev22,shabaev01,roberts22} are also listed. 
	All frequencies are given in GHz. In the table, the symbol 
	$\mu_N$ denotes the nuclear magneton.
}
	\begin{ruledtabular}
	\begin{tabular}{ll@{\hskip 0.3cm}ll@{\hskip 0.05cm}ll@{\hskip 0.5cm}ll}
		&	&\multicolumn{4}{c}{Input values} & &\\
		\cline{3-6}
		\multicolumn{2}{l}{Reference parameters} &
		\multicolumn{2}{l}{Atomic properties} &
		\multicolumn{2}{l}{Transition frequencies} &
		\multicolumn{2}{l}{Predicted parameters} \\
		\cmidrule(lr){1-2} \cmidrule(lr){3-4} \cmidrule(lr){5-6} \cmidrule(lr){7-8}\hline
		$\mu_g$~\cite{Porsev21a} & 0.366(6)~$\mu_N$
		& $T^{(0)}_{2s}$&82 268(211)~GHz/$\mu_N$    
		& $\omega_{\alpha1}$ & 28 290(5)
		&$\mu_g$ & 0.366(6)~$\mu_N$\\
		
		$\mu_m$~\cite{katori24a} & $-$0.378(8)~$\mu_N$ 
		& $T^{(0)}_{3s}$&21 347(45)~GHz/$\mu_N$  
		& $\omega_{\alpha2}$ & 31 750(5)
		&$\mu_m$ & $-$0.378(8)~$\mu_N$\\
		
		$\omega_n$~\cite{perera25a} & 2 000 161(5320) 
		&$F^{\text{ved}}_{2s}$&1 859 820(6041)~GHz/fm$^2$      
		&$\omega_{\alpha3}$ & 2 056 180(5)
		&$\omega_n$ & 2 000 161(1551) \\
		
		$\delta \langle r^{2} \rangle_{gm}$~\cite{safronova18a} & 0.0105(13)~fm$^2$
		&$F^{\text{ved}}_{3s}$&2 054 191(4990)~GHz/fm$^2$ 
		& $\omega_{\beta1}$ & 7 313(5)
		&$\delta \langle r^{2} \rangle_{gm}$& 0.0105(8)~fm$^2$\\
		
		$T_{M1}$~\cite{morgan25} & 0.84(11)~$\mu_N$ 
		& $\epsilon_{g,1s}$   &0.042(14)   
		& $\omega_{\beta2}$ & 8 266(5)
		&$T_{M1}$ & 0.84(38)~$\mu_N$ \\			
		&  & $\epsilon_{m,1s}$
		&0.053(18)    
		& $\omega_{\beta3}$
		&2 031 172(5)
		&&\\			
		&  
		&  $R_{g,2s(3s)}$ & 1.095(5)  
		&&   
		&&\\			
		&  
		& $R_{m,2s(3s)}$ & 1.094(5) 
		&&   
		&&\\
	\end{tabular}
\end{ruledtabular}
\end{table*}

With all inputs and their uncertainties specified, we determine the central 
values of the five nuclear parameters and evaluate their uncertainties using 
standard linear error propagation. For each nuclear parameter $P$, the 
uncertainty $\Delta P$ is obtained by combining the individual contributions 
in quadrature:
\begin{equation}
	\Delta P = \left[ \sum_{i} \left( \frac{\partial P}{\partial y_i} \Delta y_i \right)^2 \right]^{1/2},
	\label{un_general}
\end{equation}
where $y_i$ denotes the atomic-structure inputs and measured frequencies.
{
	Equation~(\ref{un_general}) assumes that theoretical errors are uncorrelated. To assess the impact of correlations among input uncertainties, we further perform a covariance-aware uncertainty analysis by classifying input quantities into three categories (QED-related, BW-related, and transition frequencies) and assuming full correlation within each category. As detailed in Sec.~S5 of the SM, the numerical results show that the uncertainties of \(\mu_g\) and \(\mu_m\) remain essentially unchanged, while those of \(\omega_n\), \(\delta\langle r^2\rangle_{gm}\), and \(T_{M1}\) are reduced. Thus, the independent-error assumption of Eq.~(\ref{un_general}) yields a conservative upper bound for the uncertainties of all five nuclear parameters.} The analytic expressions and numerical values of the partial derivatives for 
all five nuclear parameters are given in Sec.~{S5} of the SM 
{[Eqs.~(S46)–(S50)]}. 

Taking $\mu_g$ as an example, the corresponding partial derivatives are of order $10^{-3}$--$10^{-2}$ for the BW-related inputs $R_{g,2s(3s)}$, and $10^{-6}$--$10^{-5}$ for the transition frequencies $\omega_{\alpha1(\beta1)}$ and the uncorrected matrix elements \(T^{(0)}_{2s(3s)}\). This hierarchy indicates that the uncertainty in $\mu_g$ is dominated by the BW effect, primarily through the hydrogen-like BW correction \(\epsilon_{g,1s}\) and, to a lesser extent, the ratio factors $R_{g,2s(3s)}$, while contributions from the frequency uncertainties and the uncorrected matrix elements are negligible.

\begin{figure}[!htbp] 
	\centering 
	\includegraphics[width=0.40\textwidth]{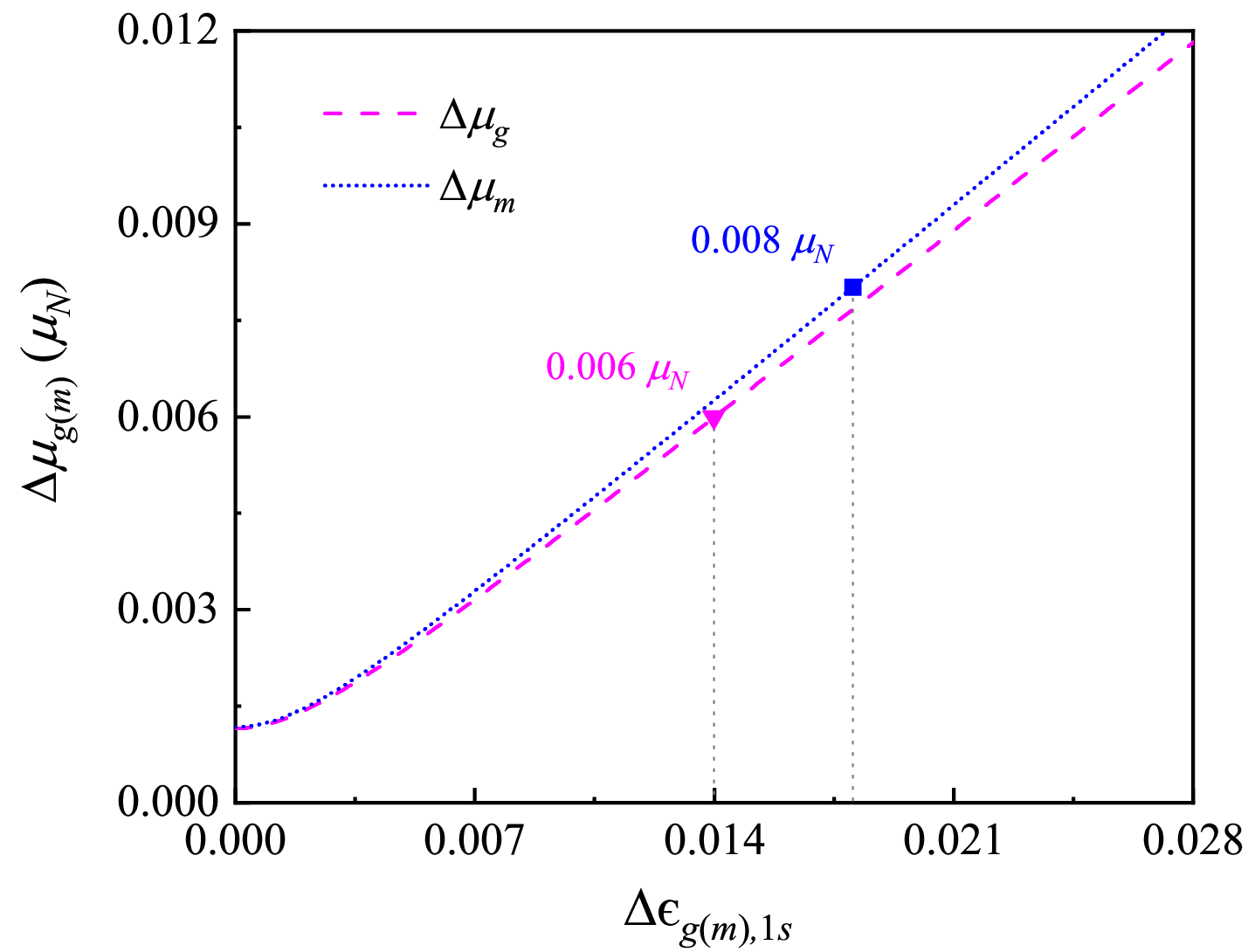}
\caption{Uncertainties in $\mu_g$ and $\mu_m$ as functions of the uncertainty 
	in the hydrogen-like BW corrections $\Delta\epsilon_{g(m),1s}$. 
	The purple triangle and blue square denote the currently predicted 
	uncertainties for the nuclear ground and isomeric states, respectively.}
	\label{fig2} 
\end{figure}

\begin{figure*}[!htbp] 
	\centering 
	\includegraphics[width=1\textwidth]{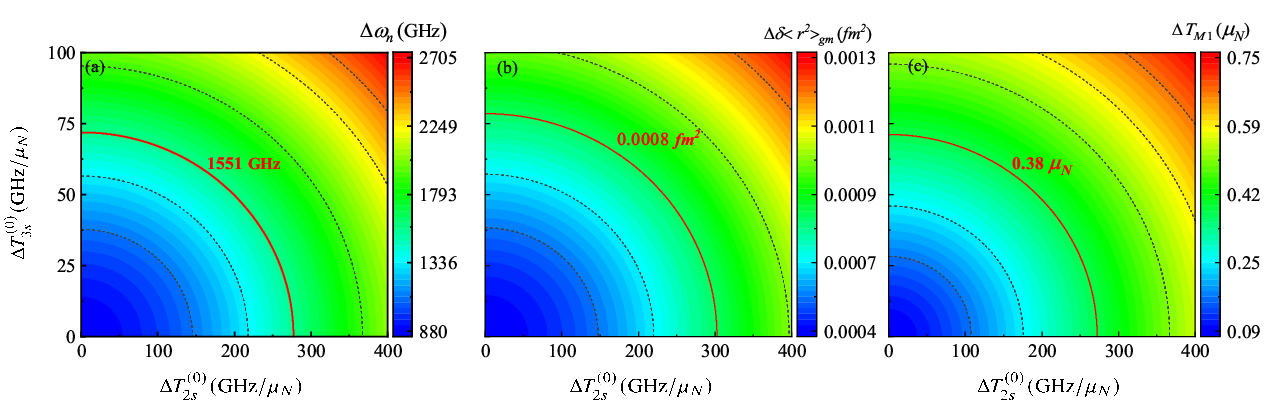}
\caption{Propagated uncertainties in (a) $\omega_n$, (b) $\delta \langle r^{2} \rangle_{gm}$, 
	and (c) $T_{M1}$ as functions of the uncertainties $\Delta T_{2s}^{(0)}$ and 
	$\Delta T_{3s}^{(0)}$ (both in GHz/$\mu_{N}$). Color contours indicate the magnitude 
	of the propagated uncertainty, while black dashed lines denote constant-value 
	contours. The red solid curves mark the estimated uncertainty of the present scheme.}
	\label{fig3} 
\end{figure*}

Evaluating these expressions with the input values listed in Table~\ref{inputs} yields the predicted uncertainties for all five nuclear parameters. Their dependence on the dominant theoretical uncertainties, particularly the BW and QED contributions to the electronic matrix elements, is illustrated in Figs.~\ref{fig2} and~\ref{fig3}.

For the nuclear magnetic moments, we obtain expected uncertainties of 
$\Delta\mu_g \simeq 6 \times 10^{-3}\,\mu_N$ and 
$\Delta\mu_m \simeq 8 \times 10^{-3}\,\mu_N$, in agreement with the current 
reference values $\mu_g = 0.366(6)\,\mu_N$~\cite{Porsev21a} and 
$\mu_m = -0.378(8)\,\mu_N$~\cite{katori24a}. These uncertainties are governed
by the electronic matrix elements, with the BW effect being the primary 
contributor. As shown in Fig.~\ref{fig2}, the uncertainties increase approximately linearly with the uncertainty of hydrogen-like BW corrections $\Delta\epsilon_{g(m),1s}$, reflecting their sensitivity to the nuclear magnetization distribution. Rather than a limitation, this sensitivity offers a unique opportunity to stimulate further research into nuclear structure.

In contrast, for $\omega_n$, $\delta \langle r^{2} \rangle_{gm}$, and $T_{M1}$, the uncertainties are dominated by QED corrections to $T_{2s}^{(0)}$ and $T_{3s}^{(0)}$. As can be seen in Fig.~\ref{fig3}, the contributions of these two matrix elements to the total uncertainty are roughly equivalent, falling within the same order of magnitude.

For the bare-nucleus transition energy, our scheme yields an uncertainty of 
1551~GHz, representing an improvement of roughly a factor of 3 compared to 
the current value of 2\,000\,161(5320)~GHz~\cite{perera25a}. The latter 
relies on an external input $\delta\langle r^{2}\rangle_{gm} = 0.0097(26)\,\mathrm{fm}^2$, 
which dominates its uncertainty. In contrast, our scheme extracts 
$\omega_n$ and $\delta\langle r^{2}\rangle_{gm}$ simultaneously, thereby 
eliminating this dominant external source of uncertainty. For 
$\delta\langle r^{2}\rangle_{gm}$, we obtain an uncertainty of 
$8\times10^{-4}\,\mathrm{fm}^2$, roughly a factor of 2 smaller than the 
current value of 0.0105(13)~$\mathrm{fm}^2$~\cite{safronova18a}. 
The extracted values of $\omega_n$ and $\delta\langle r^{2}\rangle_{gm}$ 
would provide a high-precision benchmark for testing nuclear theories.

For the nuclear reduced matrix element, we obtain 
$\Delta T_{M1} \simeq 0.38\,\mu_N$, {roughly 3.4 times larger than the uncertainty of the experimental value} $0.84(11)\,\mu_N$ obtained from radiative-lifetime measurements~\cite{morgan25}. This predicted uncertainty is dominated ($\sim 94.8\%$) by the uncorrected matrix elements 
$T_{2s(3s)}^{(0)}$, reflecting the incomplete treatment of higher-order QED contributions. The result confirms the effectiveness of our ratio-based strategy in suppressing BW-induced uncertainties, while indicating that further reduction of $\Delta T_{M1}$ will require the inclusion of higher-order QED corrections to $T_{2s}^{(0)}$ and $T_{3s}^{(0)}$.

Our numerical tests support these conclusions. Halving the QED uncertainties leaves the extracted magnetic moments nearly unchanged, while reducing the uncertainties in $\omega_n$, $\delta\langle r^2\rangle_{gm}$, and $T_{M1}$ by roughly 50\%, highlighting the importance of high-precision QED calculations. Conversely, increasing the uncertainty of the ratio factors $R_{g(m),2s(3s)}$ by an order of magnitude has little effect on the extracted parameters, demonstrating the robustness of the ratio-based approach.

\textit{Conclusion}—In summary, we have demonstrated a combined spectroscopic scheme that probes nuclear hyperfine-mixing transitions in two $^{229}$Th HCIs and enables the simultaneous determination of five key nuclear parameters. The approach exploits the strongly enhanced hyperfine-mixing regime that occurs because the large electronic energy gaps in these ions greatly exceed the energy of the first nuclear excitation.
Within this framework, the magnetic moments of the ground and isomeric nuclear states can be determined with high precision, while the bare-nucleus excitation energy, the nuclear charge-radius difference, and the nuclear transition matrix element are obtained directly from the same set of measurements. Remarkably, these results are achieved under the conservative assumption of GHz-level spectral resolution—six orders of magnitude coarser than the kHz-level resolution already demonstrated experimentally.
Even at this conservative level, the uncertainty in the bare-nucleus energy $\omega_n$ is reduced by a factor of three compared with recent measurements, while the uncertainty in the nuclear charge-radius difference $\delta\langle r^{2}\rangle_{gm}$ is reduced by a factor of two. These improvements arise not from increased spectral resolution, but from the simultaneous extraction of all nuclear parameters, which removes the need for dominant external inputs required in previous determinations.

More broadly, this work establishes a self-contained route for determining multiple nuclear parameters in systems where the first electronic level spacing far exceeds the first nuclear excitation energy. Furthermore, the theoretical framework is naturally extensible to electronic states with total angular momentum $J \ge 1$. While the present work focuses on $J=1/2$ ions, where magnetic-dipole interactions dominate, extending the scheme to higher-$J$ states would involve electric-quadrupole contributions, enabling the simultaneous and direct extraction of the nuclear quadrupole moments $Q_g$ and $Q_m$ alongside other structural parameters.

At present, the dominant uncertainty arises from atomic-structure theory, including the BW effect and QED corrections. However, continued advances in spectroscopic precision toward the kHz level will provide opportunities to probe nuclear structure in unprecedented detail. Realizing this potential would establish a definitive and self-consistent benchmark for the nuclear structure of $^{229}$Th, paving the way for the development of a nuclear optical clock and its applications in fundamental physics.

\textit{Acknowledgments}—We thank Ting-Yun Shi for valuable discussions. This work was supported by the Strategic Priority Research Program of the Chinese Academy of Sciences under Grant Nos. XDB0920000, XDB0920100, and XDB0920101; by the National Natural Science Foundation of China under Grant Nos. 12174402, 12393821, 12274417, and 12274423; and by the Chinese Academy of Sciences Project for Young Scientists in Basic Research under Grant No. YSBR-055. 
Z.-C. Y. acknowledges support from the Natural Sciences and Engineering Research Council of Canada. Some calculations were performed on the APM Theoretical Computing Cluster (APM-TCC).

\textit{Data availability}—The data supporting the findings of this Letter are available within the article and its Supplementary Material.
\appendix
\onecolumngrid
\section*{\large Supplemental Material}
\twocolumngrid
\renewcommand{\thesection}{S\arabic{section}}
\renewcommand{\thesubsection}{S\arabic{subsection}}
\renewcommand{\thetable}{S\arabic{table}}
\renewcommand{\thefigure}{S\arabic{figure}}
\setcounter{figure}{0}
\setcounter{equation}{0}   
\setcounter{table}{0}  
\renewcommand{\theequation}{S\arabic{equation}}
\subsection{Hyperfine Energy}\label{S1}
The Hamiltonian of a system consisting of nuclei and electrons is written as
\begin{eqnarray}
	H=H_n+H_e+H_{\mathrm{HFI}}, \label{hh}
\end{eqnarray}
where $H_n$, $H_e$, and $H_{\mathrm{HFI}}$ denote the nuclear Hamiltonian, the electronic Hamiltonian, and the hyperfine (electron--nucleus electromagnetic) interaction, respectively. The unperturbed operators satisfy the eigenvalue equations
\begin{eqnarray}
	H_n\ket{\gamma IM_I}=E_{\gamma I}\ket{\gamma IM_I},\label{hn}
\end{eqnarray}
and
\begin{eqnarray}
	H_e\ket{\Gamma JM_J}=E_{\Gamma J} \ket{\Gamma JM_J}.\label{he}
\end{eqnarray}
Here $\ket{\gamma I M_I}$ and $\ket{\Gamma J M_J}$ represent the nuclear and electronic eigenstates, respectively, with $\gamma$ ($\Gamma$) denoting the remaining quantum numbers.
The electronic spectrum depends on the nuclear charge distribution for a given nuclear state $x\in\{g,m\}$; the electronic energies are therefore denoted as $E_{x,\Gamma J}$ when this dependence needs to be explicit. The shorthand $E_{\Gamma J}\equiv E_{x,\Gamma J}$ is retained whenever $x$ is fixed.
The hyperfine interaction Hamiltonian in Eq.~(\ref{hh}) is written as~\cite{Schwartz55a,wang23}
\begin{eqnarray}
	H_{\mathrm{HFI}} = \sum_{kq} (-1)^{q} \mathcal{M}_{kq} T_{k-q}\,, \label{hfs1}
\end{eqnarray}
where $\mathcal{M}_{kq}$ and $T_{k-q}$ are irreducible tensor operators of rank $k$ acting on the nuclear and electronic coordinates, respectively.

In the absence of the hyperfine interaction $H_\text{HFI}$, the total system is described by product states of the form $|\gamma I M_I\rangle |\Gamma J M_J\rangle$. When $H_{\mathrm{HFI}}$ is included, it couples the nuclear and electronic angular momenta $I$ and $J$ into the conserved total angular momentum $F$, with a conserved projection $M_F$. Consequently, the appropriate basis is the coupled basis $|\gamma\Gamma I J F M_F\rangle$, which is defined as
\begin{equation}
	\ket{\gamma\Gamma IJFM_F} = \sum_{M_I, M_J} \langle IM_I, JM_J | FM_F\rangle \ket{\gamma IM_I}\ket{\Gamma JM_J}.
\end{equation}
Using the Wigner--Eckart theorem, the matrix element of $H_{\mathrm{HFI}}$ in the coupled basis reads~\cite{edmonds1957}
\begin{equation}
	\begin{aligned}
		&\langle \gamma\Gamma IJFM_F | H_{\mathrm{HFI}}
		| \gamma'\Gamma'I'J'F'M'_F \rangle 
		=	\delta_{F,F'} \delta_{M_F,M'_F}\, \\
		&\times	(-1)^{I'+F+J} \sum_k
		\left\{ \begin{matrix}
			I' & J' & F \\
			J  & I  & k \\
		\end{matrix} \right\}
		\langle \gamma I \lVert \mathcal{M}_k \rVert \gamma'I' \rangle
		\langle \Gamma J \lVert T_k \rVert \Gamma'J' \rangle .
		\label{reduce_H}
	\end{aligned}
\end{equation}

The unperturbed energy of the level $|\gamma\Gamma IJFM_F\rangle$ is $E_{\gamma I}+E_{\Gamma J}$.
Up to second order in $H_{\mathrm{HFI}}$, the hyperfine shift is
$E_{\mathrm{HFI}}^{(1)}+E_{\mathrm{HFI}}^{(2)}$, where
\begin{equation}
	E^{(1)}_{\mathrm{HFI}}=\langle \gamma\Gamma IJFM_F|H_{\mathrm{HFI}} |\gamma\Gamma IJFM_F \rangle ,  \label{delta_E1}
\end{equation}	
\begin{equation}
	E_{\mathrm{HFI}}^{(2)}=
	\sum_{\substack{\gamma'\Gamma' I'J'\\ \neq \gamma\Gamma IJ}}
	\frac{\left|\langle \gamma\Gamma IJFM_F|H_{\mathrm{HFI}}|\gamma'\Gamma' I'J'FM_F \rangle\right|^2}
	{(E_{\gamma I}+E_{\Gamma J})-(E_{\gamma'I'}+E_{\Gamma' J'})}.
	\label{delta_E2}
\end{equation}
For a fixed nuclear state $x\in\{g,m\}$, the total energy of the hyperfine level is
\begin{equation}
	E_{xF_x}=E_{\gamma_x I_x}+E_{x,\,\Gamma J}
	+E_{x,\,\mathrm{HFI}}^{(1)}+E_{x,\,\mathrm{HFI}}^{(2)}.
	\label{Ex_total}
\end{equation}

We retain only the leading magnetic-dipole ($k=1$) and electric-quadrupole ($k=2$) terms in the hyperfine interaction $H_{\mathrm{HFI}}$ of Eq.~(\ref{hfs1}). The resulting hyperfine energy shift $E_{x,\mathrm{HFI}}^{(1)}+E_{x,\mathrm{HFI}}^{(2)}$ is obtained from Eqs.~(\ref{delta_E1}) and (\ref{delta_E2}) under this truncation. Specifically, the first-order HFI energy correction can be expressed in the conventional form using hyperfine constants~\cite{trees58,Schwartz55a}:
\begin{equation}
	\begin{aligned}
		E^{(1)}_{\text{HFI}}=\frac{1}{2}AK+B\frac{\frac{3}{4}K(K+1)-I(I+1)J(J+1)}{2I(2I-1)J(2J-1)}. \label{delta_E11}
	\end{aligned}
\end{equation}	
where $K=F(F+1)-J(J+1)-I(I+1)$, and the hyperfine constants $A$ and $B$ are defined as
\begin{eqnarray}
	A = \frac{\mu}{I}\frac{1}{\sqrt{J(J+1)(2J+1)}}\langle \Gamma J\|T_1\| \Gamma J \rangle,	\label{A}
\end{eqnarray} 
\begin{eqnarray}
	B = 2Q\sqrt{\frac{J(2J-1)}{(J+1)(2J+1)(2J+3)}}\langle \Gamma J\|T_2\|\Gamma J \rangle ,	\label{B}
\end{eqnarray} 
where $\mu \equiv \mu_x$ and $Q \equiv Q_x$ are, respectively, the nuclear magnetic-dipole and electric-quadrupole moments of the nuclear state $x$. The second-order energy correction takes the form
\begin{equation}
	\begin{aligned}
		&E_{\text{HFI}}^{(2)}=\\
		&\sum_{\substack{\gamma'\Gamma' I'J'\\ \neq \gamma\Gamma IJ}} \frac{ \Big| \sum\limits_{k} \left\{ \begin{matrix}
				I' & J' & F \\
				J  & I  & k \\
			\end{matrix} \right\} 
			\left< \gamma I \lVert \mathcal{M}_k \rVert \gamma'I' \right>
			\left< \Gamma J \lVert T_k \rVert \Gamma'J' \right> \Big|^2 }
		{ \left( E_{\gamma I} - E_{\gamma' I'} \right) + \left( E_{\Gamma J} - E_{\Gamma' J'} \right)} \label{delta_E22}.
	\end{aligned}
\end{equation}	
Here the electronic energy spectrum is specific to a given nuclear state.

\subsection{Hierarchy of Energy Scales in $E_{\text{HFI}}^{(2)}$}\label{S2}

\begin{figure*}[htbp] 
	\centering 
	\includegraphics[width=0.90\textwidth]{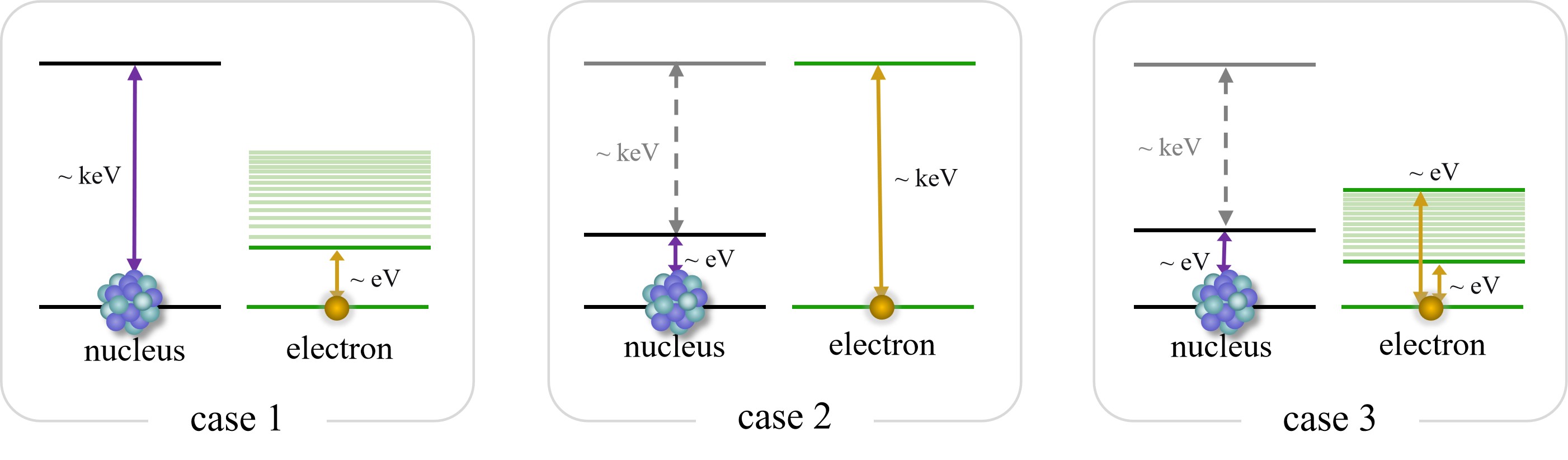}
	\caption{Schematic comparison of the nuclear and electronic level spacings
		that enter the energy denominator of the second-order perturbative correction.
		The three panels illustrate the possible hierarchies between the nuclear
		spacing $\Delta E_{\mathrm{n}}$ and the electronic spacing
		$\Delta E_{\mathrm{e}}$: (left) $\Delta E_{\mathrm{n}}\gg\Delta E_{\mathrm{e}}$
		(typical nuclei), (middle) $\Delta E_{\mathrm{n}}\ll\Delta E_{\mathrm{e}}$,
		and (right) $\Delta E_{\mathrm{n}}\sim\Delta E_{\mathrm{e}}$ (e.g., for an
		eV-scale nuclear transition such as in $^{229}$Th).}
	\label{case} 
\end{figure*}

The energy denominator in Eq.~(\ref{delta_E22}) contains both nuclear and 
electronic energy differences,
$(E_{\gamma I}-E_{\gamma' I'}) + (E_{\Gamma J}-E_{\Gamma' J'})$.
To compare their roles, it is useful to introduce the characteristic nuclear 
and electronic energy spacings,
$\Delta E_{\mathrm{n}}=|E_{\gamma I}-E_{\gamma' I'}|$ and 
$\Delta E_{\mathrm{e}}=|E_{\Gamma J}-E_{\Gamma' J'}|$, respectively.
Their relative magnitude defines three physical regimes, illustrated in 
Fig.~\ref{case}, which determine the structure of the intermediate-state 
sum in the second-order correction. We examine each regime in turn.

%


\textit{Case 1: Large Nuclear Energy Spacing (Electronic-Intermediate-State Dominated)}

The physical picture for this regime is shown in the left diagram of Fig.~\ref{case}. Characteristic nuclear energy spacings $\Delta E_{\mathrm{n}}$ are on the order of hundreds of keV, vastly exceeding the eV-scale spacings $\Delta E_{\mathrm{e}}$ of electronic levels. This large energy spacing renders contributions from intermediate states with excited nuclei negligible. The second-order correction is therefore dominated by electronic excitations, reducing Eq.~(\ref{delta_E22}) to:
\begin{equation}
	\begin{aligned}
		& E_{\text{HFI}}^{(2)}\\
		&= \sum_{\Gamma' J'} \frac{ \Big| \sum\limits_{k} \left\{ \begin{matrix}
				I  & J' & F \\
				J  & I  & k \\
			\end{matrix} \right\} 
			\left<\gamma I \lVert \mathcal{M}_k \rVert \gamma I \right>
			\left< \Gamma J \lVert T_k \rVert \Gamma'J' \right> \Big|^2 }
		{  E_{\Gamma J} - E_{\Gamma' J'}} \label{delta_E22_c1}.
	\end{aligned}
\end{equation}	
%

\textit{Case 2: Large Electronic Energy Spacing (Nuclear-Intermediate-State Dominated)}

This regime is characterized by the hierarchy $\Delta E_{\mathrm{n}} \ll \Delta E_{\mathrm{e}}$, as shown in the middle panel of Fig.~\ref{case}. The large electronic energy spacing strongly suppresses intermediate states involving electronic excitations. Therefore, the dominant contribution to the second-order HFI shift comes from intermediate states that share the same electronic configuration but differ in their nuclear state. In this limit, the energy denominator is governed by the nuclear energy difference, and Eq.~(\ref{delta_E22}) simplifies to a sum over nuclear excitations:
\begin{equation}
	\begin{aligned}
		& E_{\text{HFI}}^{(2)}\\
		&= \sum_{\gamma' I'} \frac{ \Big| \sum\limits_{k} \left\{ \begin{matrix}
				I' & J & F \\
				J  & I  & k \\
			\end{matrix} \right\} 
			\left<\gamma I \lVert \mathcal{M}_k \rVert \gamma' I' \right>
			\left<\Gamma J \lVert T_k \rVert \Gamma J \right> \Big|^2 }
		{ E_{\gamma I} - E_{\gamma' I'}} \label{delta_E22_c2}.
	\end{aligned}
\end{equation}	

This limit is precisely realized in the highly charged ions (HCIs) ${}^{229}\mathrm{Th}^{87+}$ and ${}^{229}\mathrm{Th}^{79+}$ considered in the main text. Here, the relevant nuclear energy scale is set by the ground–isomer splitting $\omega_{n} \sim 8.4~\mathrm{eV}$, while the lowest allowed electronic excitations are at $\gtrsim 7~\mathrm{keV}$, ensuring $\Delta E_{\mathrm{e}} \gg \Delta E_{\mathrm{n}}$. Higher nuclear excitations (e.g., at $\sim 29\,\mathrm{keV}$) are also off-resonant and negligible. Consequently, retaining only the dominant ground–isomer pair in Eq.~(\ref{delta_E22_c2}) yields:
\begin{widetext}
	\begin{eqnarray}
		E_{x,\,\text{HFI}}^{(2)}=  \frac{ \Big| \sum\limits_{k} \left\{ \begin{matrix}
				I_{x'} & J  & F \\
				J  & I_x  & k \\
			\end{matrix} \right\} 
			\left<\gamma_xI_x \lVert \mathcal{M}_k \rVert \gamma_{x'}I_{x'} \right>
			\left< \Gamma J \lVert T_k \rVert \Gamma J \right>_x \Big|^2 }
		{ E_{\gamma_xI_x}-E_{\gamma_{x'}I_{x'}}},\qquad (x'\neq x). \label{delta_E2g_c4}
	\end{eqnarray}	
\end{widetext}

The Kronecker $\delta$ factors in Eq.~(\ref{reduce_H}) enforce conservation of
the total angular momentum $F$ and its projection $M_F$ in each
$H_{\mathrm{HFI}}$ matrix element, thereby restricting the sum to
intermediate states with the same value of $F$. Consequently,
Eq.~(\ref{delta_E2g_c4}) receives contributions only from those
$F$ values that are compatible with both nuclear spins $I_x$ and
$I_{x'}$. The subscript $x$ on the electronic reduced matrix element
indicates that it is evaluated using the electronic wave function
corresponding to the nuclear state $x$.

\textit{Case 3: Comparable Energy Spacing }

In this regime, the nuclear and electronic energy spacings are comparable, $\Delta E_{\mathrm{n}} \sim \Delta E_{\mathrm{e}}$, as illustrated in the right panel of Fig.~\ref{case}. Both types of intermediate states must be treated on an equal footing, and the full expression of Eq.~(\ref{delta_E22}) must be used, reflecting significant nuclear-electronic coupling. This case applies to low-charge thorium ions, such as $^{229}$Th$^{3+}$~\cite{th2024}, where the nuclear isomer energy ($\sim$ 8.4 eV) lies within the range of optical electronic transitions, making $\Delta E_{\mathrm{n}}$ and $\Delta E_{\mathrm{e}}$ of similar magnitude.


\subsection{Hyperfine Transition Energies in Highly Charged $^{229}\mathrm{Th}$ Ions with $J=1/2$}\label{S3}

We now apply the general formalism of Secs.~S1 and S2 to the ions
$^{229}\mathrm{Th}^{87+}$ and $^{229}\mathrm{Th}^{79+}$ considered in the main text. The nuclear Hilbert space is restricted to the ground state ($x=g$, $I_g=5/2$) and the isomer state ($x=m$, $I_m=3/2$), while the electronic state is restricted to the $J=1/2$ ionic ground state. The resulting hyperfine levels are $F_g=2,3$ and $F_m=1,2$. 

For a $J=1/2$ electronic state, the first-order electric-quadrupole contribution vanishes. The second-order HFI shift is dominated by the nuclear ground–isomer pair with an unchanged electronic configuration. Accordingly, Eq.~(\ref{delta_E2g_c4}) applies. Starting from the total energy in Eq.~(\ref{Ex_total}) and inserting the first- and second-order HFI shifts under these approximations, we obtain 
\begin{equation}
	\begin{aligned}
		E_{xF_x}
		&= E_{\gamma_x I_x}+E_{x,\Gamma J}
		+ C_1(F_x,I_x)\,\mu_x\,T_x \\
		&\quad
		+ C_2(F_x,I_x,I_{x'})\,
		\frac{T_{M1}^2\,T_x^2}{\omega_{xx'}} ,
		\qquad (x'\neq x).
	\end{aligned}
	\label{key}
\end{equation}
where $\omega_{xx'}= E_{\gamma_x I_x}-E_{\gamma_{x'} I_{x'}} $ is the bare nuclear transition energy. 
The coefficients $C_1(F_x,I_x)$ and $C_2(F_x,I_x,I_{x'})$ are defined as
\begin{equation}
	\begin{aligned}
		C_1(F_x,I_x)
		&=\frac{F_x(F_x+1)-J(J+1)-I_x(I_x+1)}
		{2I_x\sqrt{J(J+1)(2J+1)}} ,
	\end{aligned}
	\label{c1}
\end{equation}
\begin{equation}
	\begin{aligned}
		C_2(F_x,I_x,I_{x'})
		=
		\begin{Bmatrix}
			I_{x'} & J & F_x\\
			J      & I_x & 1
		\end{Bmatrix}^{2}.
	\end{aligned}
	\label{c2}
\end{equation}
$T_x = \langle \Gamma J \| T_1 \| \Gamma J \rangle_x $ and $T_{M1}= \langle \gamma_g I_g \| \mathcal{M}_1 \| \gamma_m I_m \rangle$ denote the electronic and nuclear magnetic-dipole transition matrix elements, respectively. 

The hyperfine transition frequency between two levels is defined as $\omega_{F_x\rightarrow F_{x'}} \equiv E_{xF_x}-E_{x'F_{x'}}$. Using Eq.~(\ref{key}), we derive the explicit expression
\begin{widetext}
	\begin{equation}
		\begin{aligned}
			\omega_{F_x\rightarrow F_{x'}}
			&=\omega_{xx'} +\left(E_{x,\Gamma J}-E_{x',\Gamma J}\right)
			+\Big[C_1(F_x,I_x)\mu_xT_x-C_1(F_{x'},I_{x'})\mu_{x'}T_{x'}
			\Big] \\
			&\quad+\Bigg[
			C_2(F_x,I_x,I_{x''})\frac{T_{M1}^2T_x^2}{\omega_{xx''}}
			-
			C_2(F_{x'},I_{x'},I_{x'''})\frac{T_{M1}^2T_{x'}^2}{\omega_{x'x'''}}
			\Bigg].
		\end{aligned}
		\label{eq:nu_diff_x}
	\end{equation}
\end{widetext}
%

The second term in Eq.~(\ref{eq:nu_diff_x}) is the electronic field shift (FS) arising from the change in the nuclear charge distribution between states $x$ and $x'$. For the ground-isomer pair ($x=m$ and $x'=g$), it can be expressed as~\cite{torbohm85a}
\begin{equation}
	E_{\mathrm{fs}}=E_{m,\Gamma J}-E_{g,\Gamma J}
	=\sum_{n \geq 0, \, \mathrm{even}} F_n \delta \langle r^{n+2} \rangle_{gm}, \label{field-shift0}
\end{equation}
where $F_n$ are the field shift factors and $\delta \langle r^{n+2} \rangle_{gm}$ denotes the difference in the $(n+2)$-th moment of the nuclear charge radius between the nuclear ground state and the isomeric state. To account for the varying electron density (VED) inside the nucleus, we follow Ref.~\cite{ekman19a} and group higher-order moments into an effective factor. Retaining the first four terms yields
\begin{widetext}
	\begin{equation}
		\begin{aligned}
			E_{\mathrm{fs}}
			=\left[
			F_0
			+F_2\frac{\delta\langle r^{4} \rangle_{gm}}{\delta\langle r^{2} \rangle_{gm}}
			+F_4\frac{\delta\langle r^{6} \rangle_{gm}}{\delta\langle r^{2} \rangle_{gm}}
			+F_6\frac{\delta\langle r^{8} \rangle_{gm}}{\delta\langle r^{2} \rangle_{gm}}
			\right]\delta\langle r^{2} \rangle_{gm}
			\equiv F^{\mathrm{ved}}\,\delta\langle r^{2} \rangle_{gm},
		\end{aligned}
		\label{field-shift}
	\end{equation}
\end{widetext}
which improves upon the leading-order approximation $E_{\mathrm{fs}}\approx F_0\,\delta\langle r^{2}\rangle_{gm}$ used in previous work~\cite{zheng26}. Here $F^{\mathrm{ved}}$ is calculated using the RIS4 package~\cite{ekman19a} with a Fermi distribution for the nuclear charge density~\cite{parpia92a}, and we adopt $\delta \langle r^{2} \rangle_{gm} = 0.0105(13)$ fm$^2$ from Ref.~\cite{safronova18a}.

{\subsection{Radiative Lifetimes of Hyperfine States}}
\label{sec:S4_lifetime}

{In the $^{229}\mathrm{Th}$ nucleus, the spontaneous decay of the nuclear isomeric state is dominated by magnetic-dipole ($M1$) transitions~\cite{tiedau24,Porsev10}. Under the influence of nuclear hyperfine mixing (NHM), we therefore focus on the $M1$ transition matrix elements and the resulting radiative lifetimes of the excited hyperfine states for the $^{229}\mathrm{Th}^{87+}$ and $^{229}\mathrm{Th}^{79+}$ ions. }

{For a transition from an initial hyperfine state $i$ to a final hyperfine state $f$, the $M1$ spontaneous-emission rate is given by~\cite{shabaev22}}
\begin{equation}
	A_{i\to f}^{M1}
	=
	\frac{1}{4\pi}
	\frac{4}{3}
	\nu_{if}^{3}
	\frac{
		\left|
		M_{i\to f}
		\right|^2
	}{
		2F_i+1
	},
	\label{eq:S4_M1_rate}
\end{equation}
{where \(F_i\) is the total angular momentum of the initial state,
	\(\nu_{if}\) is the transition frequency in natural units, and}
\begin{equation}
	M_{i\to f}
	=
	\left\langle f
	\left\|
	\mathcal{M}_1^{\rm rad}
	\right\|
	i
	\right\rangle
\end{equation}
{is the reduced $M1$ matrix element. When NHM is relevant, the affected state is replaced by its perturbed counterpart, \(\ket{i}\to\ket{i}_{+}\) or
	\(\ket{f}\to\ket{f}_{+}\).}

{For numerical evaluation, the transition frequency $\omega_{if}$ is taken in GHz and the reduced matrix element in units of \(\mu_N\). Eq.~(\ref{eq:S4_M1_rate}) then becomes}
\begin{equation}
	A_{i\to f}^{M1}=2.96932426\times10^{-22}\,\omega_{if}^{3}\frac{\left|M_{i\to f}/\mu_N\right|^2}{2F_i+1}
	\quad {\rm s}^{-1}.
	\label{eq:S4_M1_rate_numeric}
\end{equation}

{The total radiative $M1$ operator is}
\begin{equation}
	\mathcal{M}_1^{\rm rad}
	=
	\mathcal{M}_1+\mathcal{M}_1^{(e)} ,
	\label{eq:S4_total_M1_operator}
\end{equation}
{where \(\mathcal{M}_1\) is the nuclear $M1$ transition operator introduced in Eq.~(\ref{hfs1}), and \(\mathcal{M}_1^{(e)}\) is the electronic $M1$ transition operator, which should be distinguished from the electronic tensor operator $T_1$ appearing in the hyperfine interaction. For a valence state $v$, we define}
\begin{align}
	T_{v}&\equiv\left\langle v\left\|T_1\right\|v\right\rangle ,\\
	\mu_v^{(e)}&\equiv\left\langle v\left\|\mathcal{M}_1^{(e)}\right\|v\right\rangle ,\\
	T_{M1}&\equiv\left\langle \gamma_g I_g\left\|\mathcal{M}_1\right\|\gamma_mI_m\right\rangle .
\end{align}
{Here $T_v$ is the reduced matrix element of the electronic hyperfine tensor operator entering the NHM coefficient, while $\mu_v^{(e)}$ is the reduced matrix element of the electronic radiative $M1$ operator.}

{For the $J=1/2$ hyperfine states, only the  $F=2$ states associated with the nuclear ground and isomeric states are affected by NHM. Introducing}
\begin{equation}
	\eta_v
	=
	\frac{T_{v}T_{M1}}{E_{\gamma_mI_m}-E_{\gamma_gI_g}},
	\label{eq:S4_eta}
\end{equation}
{and using \(I_g=5/2\), \(I_m=3/2\), and \(J=1/2\), the $M1$ transition matrix elements between various hyperfine states are given by:}
\begingroup
\small
\begin{equation}
	\begin{aligned}
	M_{|I_m,F=2\rangle_+ \to |I_g,F=2\rangle_+}
	&\equiv	{}_{+}\!\left\langle I_g,v,F=2\left\|\mathcal{M}_1^{\rm rad}\right\|I_m,v,F=2\right\rangle_{+}
	\nonumber\\
	&=-\frac{\sqrt{3}}{6}\,T_{M1}-\frac{5\sqrt{3}}{18}\,\eta_v\,\mu_v^{(e)},
	\label{eq:S4_M_mF2_gF2}
\end{aligned}
\end{equation}
\begin{equation}
	\begin{aligned}		
	M_{|I_m,F=2\rangle_+\to |I_g,F=3\rangle}
	&\equiv\left\langle I_g,v,F=3\left\|\mathcal{M}_1^{\rm rad}	\right\|
	I_m,v,F=2\right\rangle_{+}
	\nonumber\\
	&=\frac{\sqrt{42}}{6}\,T_{M1}+\frac{\sqrt{42}}{18}\,\eta_v\,\mu_v^{(e)},
	\label{eq:S4_M_mF2_gF3}
\end{aligned}
\end{equation}
	\begin{equation}
		\begin{aligned}
	M_{|I_m,F=1\rangle \to |I_g,F=2\rangle_+}&\equiv
	{}_{+}\!\left\langle I_g,v,F=2\left\|\mathcal{M}_1^{\rm rad}
	\right\|
	I_m,v,F=1\right\rangle
	\nonumber\\
	&=\frac{\sqrt{3}}{2}\,T_{M1}-\frac{\sqrt{3}}{6}\,
	\eta_v\,\mu_v^{(e)},
	\label{eq:S4_M_mF1_gF2}
\end{aligned}
\end{equation}
\begin{equation}
	\begin{aligned}
	M_{|I_m,F=1\rangle \to |I_m,F=2\rangle_+}
	&\equiv{}_{+}\!\left\langle I_m,v,F=2\left\|\mathcal{M}_1^{\rm rad}
	\right\|I_m,v,F=1\right\rangle
	\nonumber\\
	&=\frac{\sqrt{5}}{2}\,\mu_v^{(e)}+\frac{\sqrt{5}}{10}\,\eta_v\,T_{M1},
	\label{eq:S4_M_mF1_mF2}
\end{aligned}
\end{equation}
	\begin{equation}
		\begin{aligned}
	M_{|I_g,F=3\rangle \to |I_g,F=2\rangle_+}
	&\equiv{}_{+}\!\left\langle I_g,v,F=2\left\|\mathcal{M}_1^{\rm rad}
	\right\|I_g,v,F=3\right\rangle
	\nonumber\\
	&=-\frac{\sqrt{70}}{6}\,\mu_v^{(e)}+\frac{\sqrt{70}}{30}\,\eta_vT_{M1}.
	\label{eq:S4_M_gF3_gF2}
\end{aligned}
\end{equation}
\endgroup
{In our calculations, we take $v = 2s_{1/2}$ for the $^{229}\mathrm{Th}^{87+}$ ion and $v = 3s_{1/2}$ for $^{229}\mathrm{Th}^{79+}$ ion, and use the data from Table~I of the main text together with the following electronic matrix elements:}
\begin{align}
	\mu_{2s}^{(e)}
	&=
	\left\langle 2s_{1/2}
	\left\|
	\mathcal{M}_1^{(e)}
	\right\|
	2s_{1/2}
	\right\rangle
	=
	-2.3520\,\mu_B ,
	\\
	\mu_{3s}^{(e)}
	&=
	\left\langle 3s_{1/2}
	\left\|
	\mathcal{M}_1^{(e)}
	\right\|
	3s_{1/2}
	\right\rangle
	=
	-2.4903\,\mu_B .
\end{align}
{These values are converted to $\mu_N$ when evaluating Eq.~(\ref{eq:S4_M1_rate_numeric}). The resulting matrix elements, transition frequencies, $M1$ transition rates, and the corresponding lifetimes are listed in Table~\ref{tab:S4_transition_rates}.}

{The total radiative decay rate of an excited hyperfine state is obtained by summing over all energetically allowed downward $M1$ channels,}
\begin{equation}
	A_{\rm tot}(i)
	=
	\sum_f A_{i\to f}.
\end{equation}
{The lifetime and natural linewidth are then given by}
\begin{equation}
	\tau_i
	=
	\frac{1}{A_{\rm tot}(i)},
	\qquad
	\Delta\nu_{\rm nat}
	=
	\frac{A_{\rm tot}(i)}{2\pi}.
\end{equation}
{The results are summarized in Table~\ref{tab:S4_total_lifetimes}.}

{The transitions listed in Table~\ref{tab:S4_transition_rates} include both intramanifold hyperfine transitions and intermanifold transitions. The intramanifold hyperfine transitions are dominated by the electronic radiative $M1$ matrix element. For the intermanifold transitions, which would otherwise be governed mainly by the nuclear $M1$ matrix element, NHM introduces additional electronic radiative $M1$ contributions. This makes the corresponding reduced matrix elements significantly larger than the bare nuclear value \(T_{M1}=0.84(11)\,\mu_N\)~\cite{morgan25}. For the intermanifold hyperfine transitions, our calculated lifetimes lie within the range reported by Shabaev et al.~\cite{shabaev22}. However, mainly owing to the use of different $\omega_{if}$, our lifetimes are shorter than those of Shabaev et al~\cite{shabaev22} for the intramanifold hyperfine transitions. The total lifetimes summarized in Table~\ref{tab:S4_total_lifetimes} show that the isomeric hyperfine states are shortened to less than 1 s in \(^{229}\mathrm{Th}^{87+}\) and to a few seconds in \(^{229}\mathrm{Th}^{79+}\). The corresponding natural linewidths are all below \(1~{\rm Hz}\), and thus do not limit the GHz-level spectroscopic precision assumed in the present work.}
\\

\begin{table*}[!htbp]
	\tabcolsep=0.0005cm
	\renewcommand\arraystretch{1.3}
   {\caption{
			Transition frequencies, reduced $M1$ matrix elements, transition
			rates, and the corresponding lifetimes of the \(^{229}\mathrm{Th}^{87+}\)
			and \(^{229}\mathrm{Th}^{79+}\) ions. The present results use
			\(T_{M1}=0.84(11)\,\mu_N\)~\cite{morgan25}, corresponding to
			the nuclear $M1$ reduced transition probability \(B(M1)=T_{M1}^{2}/30=0.0235(62)\) W.u. The last column lists the
			lifetimes of the \(^{229}\mathrm{Th}^{87+}\) ion from Ref.~\cite{shabaev22}
			for a range of $B(M1)$ values from 0.005 to \(0.048\) W.u.
		}
		\label{tab:S4_transition_rates}}
	\begin{ruledtabular}
		\begin{tabular}{lccccc}
			\multicolumn{1}{c}{Transition} & \(\omega_{i f}\) (GHz) &
			\(|M_{i\to f}|\) \((\mu_N)\) & \(A_{i\to f}\) (s\(^{-1}\)) &
			\(1/A_{i\to f}\) (s) & \(1/A_{i\to f}\) (s)~\cite{shabaev22} \\
			\hline
			\multicolumn{6}{c}{\(^{229}\mathrm{Th}^{87+}\)}\\
			$|I_m,F_i=2\rangle_+ \rightarrow |I_g,F_f=2\rangle_+$& \(2\,024\,430(5)\) & \(71.545\)& \(2.522\) & \(0.397\) & \(0.229 - 2.20\) \\
			$|I_m,F_i=2\rangle_+\rightarrow |I_g,F_f=3\rangle$& \(1\,996\,140(5)\) & \(52.814\)& \(1.318\) & \(0.759\) & \(0.408 - 3.92\) \\
			$|I_m,F_i=1\rangle \rightarrow |I_g,F_f=2\rangle_+$& \(2\,056\,180(5)\) & \(43.800\)& \(1.651\) & \(0.606\) & \(0.375 - 3.60\) \\
			$|I_m,F_i=1\rangle \rightarrow |I_m,F_f=2\rangle_+$& \(31\,750(5)\) & \(4828.370\)& \(7.385\times10^{-2}\) & \(13.541\) & \(14.8 - 15.2\) \\		
			$|I_g,F_i=3\rangle \rightarrow |I_g,F_f=2\rangle_+$& \(28\,290(5)\) & \(6022.051\)& \(3.483\times10^{-2}\) & \(28.711\) & \(30.8 - 31.6\) \\
			\hline
			\multicolumn{6}{c}{\(^{229}\mathrm{Th}^{79+}\)}\\
			$|I_m,F_i=2\rangle_+ \rightarrow |I_g,F_f=2\rangle_+$& \(2\,022\,906(5)\) & \(19.480\)& \(0.187\) & \(5.348\) & -- \\
			$|I_m,F_i=2\rangle_+\rightarrow |I_g,F_f=3\rangle$& \(2\,015\,633(5)\) & \(13.852\)& \(9.331\times10^{-2}\) & \(10.717\) & -- \\
			$|I_m,F_i=1\rangle \rightarrow |I_g,F_f=2\rangle_+$& \(2\,031\,172(5)\) & \(12.561\)& \(0.131\) & \(7.634\) & -- \\
			$|I_m,F_i=1\rangle \rightarrow |I_m,F_f=2\rangle_+$& \(8\,226(5)\) & \(5112.288\)& \(1.440\times10^{-3}\) & \(694.444\) & -- \\
			$|I_g,F_i=3\rangle \rightarrow |I_g,F_f=2\rangle_+$& \(7\,313(5)\) & \(6376.148\)& \(6.745\times10^{-4}\) & \(1482.580\) & --
		\end{tabular}
	\end{ruledtabular}
\end{table*}
\begin{table}[!htbp]
	\tabcolsep=0.0005cm
	\renewcommand\arraystretch{1.3}
	{	\caption{
			The total $M1$ decay rates, radiative lifetimes, and natural linewidths of the each excited hyperfine states for the $^{229}$Th$^{87+}$ and $^{229}$Th$^{79+}$ ions. }
		\label{tab:S4_total_lifetimes}}
	\begin{ruledtabular}
		\begin{tabular}{lccc}
			\multicolumn{1}{c}{State} & \(A_{\rm tot}\) (s\(^{-1}\)) &\(\tau\) (s) & \(\Delta\nu_{\rm nat}\) (Hz)  \\
			\hline
			\multicolumn{4}{c}{\(^{229}\mathrm{Th}^{87+}\)}\\
			$|I_m,F=1\rangle$   & \(1.725\) & \(0.58\)   & \(0.28\) \\
			$|I_m,F=2\rangle_+$ & \(3.840\) & \(0.26\)   & \(0.61\)\\
			$|I_g,F=3\rangle$   & \(3.483\times10^{-2}\) & \(28.71\) &\(5.54\times10^{-3}\)  \\
			\hline
			\multicolumn{4}{c}{\(^{229}\mathrm{Th}^{79+}\)}\\
			$|I_m,F=1\rangle$   & \(0.132\) & \(7.58\)   & \(2.10\times10^{-2}\)  \\
			$|I_m,F=2\rangle_+$ & \(0.280\) & \(3.57\)   & \(4.46\times10^{-2}\) \\
			$|I_g,F=3\rangle$   & \(6.745\times10^{-4}\) & \(1482.58\) &\(1.07\times10^{-4}\) 
		\end{tabular}
	\end{ruledtabular}
\end{table}

\subsection{Derivation of the Nuclear Parameter Uncertainty Formula}\label{S4}

We first evaluate the electronic $M1$ matrix elements at a pointlike magnetization of the nucleus; the ground-isomer charge-radius difference gives a negligible correction, yielding the nuclear-state-independent values $T^{(0)}_{2s}$ and $T^{(0)}_{3s}$. These quantities are obtained from MCDHF calculations that include electron
correlation, the Breit interaction, and QED corrections. The extrapolated values of \(T^{(0)}_{2s(3s)}\) are listed in the last row of Table~\ref{active_space}.
The theoretical uncertainty is dominated by higher-order QED effects; as an estimate, we take one half of the computed QED correction.

Table~\ref{active_space} also lists the field-shift factors \(F_{2s}^{\mathrm{ved}}\) and
\(F_{3s}^{\mathrm{ved}}\) for $^{229}$Th$^{87+}$ and $^{229}$Th$^{79+}$ ions, obtained within the same MCDHF framework. Their uncertainties are likewise governed by higher-order QED contributions and are evaluated using the same prescription.
\begin{table*}[!htbp]
	\tabcolsep=0.0005cm
	\renewcommand\arraystretch{1.5}
	\caption{\label{active_space}
		Values of the electronic matrix elements (in GHz/$\mu_N$) and the field-shift
		factors $F^\text{ved}_{2s}$ and $F^\text{ved}_{3s}$ (in GHz/fm$^2$) obtained
		with different active orbital spaces. “Breit” and “QED” indicate the inclusion
		of the Breit interaction and QED corrections, respectively, while “Extrap.”
		denotes an extrapolated value. Numbers in parentheses represent the
		uncertainty arising from higher-order QED corrections.}
	\begin{ruledtabular}
		\begin{tabular}{ccllcll}
			&\multicolumn{3}{c}{$^{229}$Th$^{87+}$}&\multicolumn{3}{c}{$^{229}$Th$^{79+}$}\\
			\cline{2-4}\cline{5-7}
			\multicolumn{1}{c}{Layer} & \multicolumn{1}{c}{Active orbitals ($n_{\rm max}\,\ell$)} & \multicolumn{1}{c}{$T_{2s}^{(0)}$} & \multicolumn{1}{c}{$F_{2s}^{\text{ved}}$} & \multicolumn{1}{c}{Active orbitals ($n_{\rm max}\,\ell$)} & \multicolumn{1}{c}{$T_{3s}^{(0)}$} & \multicolumn{1}{c}{$F_{3s}^\text{ved}$} 	\\
			\hline			
			DHF &$1s,2s$               & 80 965.20  & 1 858 436 & $3s,2p$             & 20 912.63 &2 055 845\\  	
			1 & $3s,2p,3d,4f,5g,6h$    & 81 804.95  & 1 857 987 & $4s,3p,3d,4f,5g,6h$ & 21 186.12 &2 055 829\\
			2 & $4s,3p,4d,5f,6g,6h$    & 81 805.79  & 1 857 987 & $5s,4p,4d,5f,6g,7h$ & 21 213.28 &2 055 853\\
			3 & $5s,4p,5d,6f,6g,6h$    & 81 734.35  & 1 858 045 & $6s,5p,5d,6f,7g,8h$ & 21 239.21 &2 055 856\\
			4 & $6s,5p,6d,6f,6g,6h$    & 81 754.56  & 1 858 000 & $7s,6p,6d,7f,8g,9h$ & 21 225.04 &2 055 863\\	
			5 & $7s,6p,7d,6f,6g,6h$    & 81 747.10  & 1 858 037 & $8s,7p,7d,8f,9g,9h$ & 21 230.80 &2 055 866\\
			6 & $8s,7p,8d,6f,6g,6h$    & 81 749.90  & 1 858 012 & $9s,8p,8d,9f,9g,9h$ & 21 227.79 &2 055 868\\
			& Layer 5+Breit            & 81 845.91  & 1 847 738 &Layer 5+Breit        & 21 256.72 &2 044 211\\
			&Layer 5+Breit+QED                     & 82 267.59  & 1 859 820 &Layer 5+Breit+QED                & 21 346.79 &2 054 191\\
			& Extrap.                  &  82 268(211)  & 1 859 820(6041) & Extrap.    & 21 347(45)&2 054 191(4990)\\
		\end{tabular}
	\end{ruledtabular}
\end{table*}

To incorporate the BW effect, we relate the BW corrections for Li-like (\(2s\)) $^{229}$Th$^{87+}$ and Na-like (\(3s\)) $^{229}$Th$^{79+}$ 
ions to those for the corresponding H-like (\(1s\)) $^{229}$Th$^{89+}$ ion by introducing the ratio factors:
\begin{align}
	R_{g,2s} &= \frac{\epsilon_{g,2s}}{\epsilon_{g,1s}}, &
	R_{m,2s} &= \frac{\epsilon_{m,2s}}{\epsilon_{m,1s}}, \label{Rg2sRm2s}\\
	R_{g,3s} &= \frac{\epsilon_{g,3s}}{\epsilon_{g,1s}}, &
	R_{m,3s} &= \frac{\epsilon_{m,3s}}{\epsilon_{m,1s}}. \label{Rg3sRm3s}
\end{align}
Here, \(\epsilon_{g,1s}\) and \(\epsilon_{m,1s}\) are the BW corrections for the $^{229}$Th$^{89+}$ ion in the
nuclear ground and isomeric states, respectively, while \(\epsilon_{g,2s}\) (\(\epsilon_{m,2s}\)) and
\(\epsilon_{g,3s}\) (\(\epsilon_{m,3s}\)) are the corresponding corrections for the Li-like (\(2s\)) and Na-like (\(3s\)) $^{229}$Th ions. Using the ratio factors in Eqs.~\eqref{Rg2sRm2s}--\eqref{Rg3sRm3s}, the
BW-inclusive $M1$ matrix elements are obtained from the
uncorrected ones as:
\begin{align}
	T_{g,2s} &= \bigl(1-R_{g,2s}\epsilon_{g,1s}\bigr)\,T^{(0)}_{2s}, \label{Tg2s}\\
	T_{m,2s} &= \bigl(1-R_{m,2s}\epsilon_{m,1s}\bigr)\,T^{(0)}_{2s}, \label{Tm2s}\\
	T_{g,3s} &= \bigl(1-R_{g,3s}\epsilon_{g,1s}\bigr)\,T^{(0)}_{3s}, \label{Tg3s}\\
	T_{m,3s} &= \bigl(1-R_{m,3s}\epsilon_{m,1s}\bigr)\,T^{(0)}_{3s}. \label{Tm3s}
\end{align}

The BW corrections for the $1s$ and $2s$ states are tabulated in Table~I of the Supplemental Material of Ref.~\cite{shabaev22}, from which we take the central values of $R_{g,2s}$ and $R_{m,2s}$. The ratio factors can be determined with substantially higher accuracy than the absolute BW contributions, owing to their weak sensitivity to the assumed nuclear magnetization model. This insensitivity reflects the near proportionality of $ns$ Dirac wave functions inside the nucleus~\cite{shabaev01,roberts22}. We therefore conservatively assign an absolute uncertainty of $0.005$ to both ratios and list the adopted values in Table~I of the main text.

For the Na-like ($3s$) Th ion, dedicated BW-ratio data are currently unavailable, and we have not yet performed an explicit calculation of $R_{g(m),3s}$. However, for $ns$ states the relative BW correction depends only weakly on the principal quantum number $n$, as all the electronic wave functions in the nuclear region are nearly proportional to each other for fixed $\kappa$ ($\kappa = (L-J)(2J+1)$)~\cite{bohr50,ginges18,roberts22}.
We therefore adopt $R_{g,3s}\approx R_{g,2s}$ and
$R_{m,3s}\approx R_{m,2s}$ as reasonable proxies in the absence of
dedicated Na-like $^{229}$Th data. {The impact of this approximation on the extracted uncertainties is assessed after the uncertainty-propagation analysis.}

To assess the achievable precision of the nuclear parameters extracted from
Eqs.~(8)--(12) of the main text, one would ideally require experimental values
for all six transition frequencies $\omega_{\alpha i}$ and $\omega_{\beta i}$. Since such measurements are not yet available, we generate representative central values by inserting reference nuclear parameters and atomic properties into Eqs.~(1)--(6) of the main text. These values serve only as reference points for the subsequent analysis. In this step, we use the proxy \(R_{g,3s}=R_{g,2s}\) and \(R_{m,3s}=R_{m,2s}\) to obtain the BW-corrected \(3s\)
matrix elements for the $^{229}$Th$^{79+}$ ion, justified by the weak $n$-dependence of the relative BW correction for $ns$ states~\cite{bohr50,ginges18,roberts22}. This approximation is used solely to define the reference line centers; once experimental frequencies become available, they enter directly as inputs, and the auxiliary central values are no longer needed.

Using the extrapolated atomic-structure values in Table~\ref{active_space}, together with the reference nuclear parameters and BW-corrected inputs listed in Table~I of the main text, we evaluate the six hyperfine transition line centers and list them in the third column in Table~I. The measurement uncertainty for these frequencies is assumed to be 5 GHz, comparable to the precision demonstrated for highly charged Bi$^{80+}$ ions~\cite{ullmann17}.

{Neglecting correlations among the input uncertainties}, we apply the standard linear uncertainty propagation formula (Eq.~(15) of the main text) to derive analytic expressions for the uncertainties of the extracted nuclear parameters. These expressions depend on the assumed experimental uncertainties of the six transition frequencies and the theoretical uncertainties of \(R\), \(\epsilon_{1s}\), \(T^{(0)}\), and \(F^{\mathrm{ved}}\). The expanded form of the uncertainties for each nuclear parameter can be written as:
\begin{widetext}
	\begin{equation}
		\begin{aligned}
			\Delta \mu_g &= \biggl[ (-5.7042\times 10^{-3})^2(\Delta R_{g,2s})^2 + (2.1817\times 10^{-2})^2(\Delta R_{g,3s})^2 + (0.4201)^2(\Delta \epsilon_{g,1s})^2  + (1.5750\times 10^{-6})^2(\Delta T^{(0)}_{2s})^2 \\
			&\quad+ (-2.3215\times 10^{-5})^2(\Delta T^{(0)}_{3s})^2 + (-4.5567\times 10^{-6})^2(\Delta \omega_{\alpha 1})^2  + (6.7677\times 10^{-5})^2(\Delta \omega_{\beta 1})^2 \biggr]^{1/2},
		\end{aligned}
		\label{un_mug}
	\end{equation}
	\begin{equation}
		\begin{aligned}
			\Delta \mu_m = \biggl[ &(7.3861\times 10^{-3})^2(\Delta R_{m,2s})^2 + (-2.8653\times 10^{-2})^2(\Delta R_{m,3s})^2 + (-0.4390)^2(\Delta \epsilon_{m,1s})^2 \\
			& + (-1.5958\times 10^{-6})^2(\Delta T^{(0)}_{2s})^2 + (2.3857\times 10^{-5})^2(\Delta T^{(0)}_{3s})^2 + (4.1532\times 10^{-6})^2(\Delta \omega_{\alpha 2})^2 \\
			& + (-6.1684\times 10^{-5})^2(\Delta \omega_{\beta 2})^2 \biggr]^{1/2},
		\end{aligned}
		\label{un_mum}
	\end{equation}
	\begin{equation}
		\begin{aligned}
			\Delta\omega_n = \biggl[ & (-1.6873\times 10^{4})^2(\Delta R_{g,2s})^2 + (272.96)^2(\Delta R_{m,2s})^2 + (1.6673\times 10^{4})^2(\Delta R_{g,3s})^2 + (-16.639)^2(\Delta R_{m,3s})^2 \\
			& + (-5229.0)^2(\Delta \epsilon_{g,1s})^2 + (5290.8)^2(\Delta \epsilon_{m,1s})^2 + (4.5998)^2(\Delta T^{(0)}_{2s})^2 + (-17.727)^2(\Delta T^{(0)}_{3s})^2 \\
			& + (-19.581)^2(\Delta \omega_{\alpha 1})^2 + (-6.6053)^2(\Delta \omega_{\alpha2})^2 + (10.568)^2(\Delta \omega_{\alpha3})^2 + (57.285)^2(\Delta \omega_{\beta1})^2 \\
			& + (5.9803)^2(\Delta \omega_{\beta2})^2 + (-9.5684)^2(\Delta \omega_{\beta3})^2 + (-0.1110)^2 (\Delta F^\text{ved}_{2s})^2 + (0.1005)^2 (\Delta F^\text{ved}_{3s})^2 \biggr]^{1/2},
		\end{aligned}
		\label{un_wn}
	\end{equation}
	\begin{equation}
		\begin{aligned}
			\Delta\delta\langle r^{2} \rangle_{gm} = \biggl[ &(8.1585\times 10^{-3})^2(\Delta R_{g,2s})^2 + (-1.3288\times 10^{-4})^2(\Delta R_{m,2s})^2 + (-8.0616\times 10^{-3})^2(\Delta R_{g,3s})^2 \\
			& + (8.9467\times 10^{-6})^2(\Delta R_{m,3s})^2 + (2.5283\times 10^{-3})^2(\Delta \epsilon_{g,1s})^2 + (-2.5581\times 10^{-3})^2(\Delta \epsilon_{m,1s})^2 \\
			& + (-2.2239\times 10^{-6})^2(\Delta T^{(0)}_{2s})^2 + (8.5706\times 10^{-6})^2(\Delta T^{(0)}_{3s})^2 + (9.4879\times 10^{-6})^2(\Delta \omega_{\alpha 1})^2 \\
			& + (3.2155\times 10^{-6})^2(\Delta \omega_{\alpha2})^2 + (-5.1448\times 10^{-6})^2(\Delta \omega_{\alpha3})^2 + (-2.8000\times 10^{-5})^2(\Delta \omega_{\beta1})^2 \\
			& + (-3.2155\times 10^{-6})^2(\Delta \omega_{\beta2})^2 + (5.1448\times 10^{-6})^2(\Delta \omega_{\beta3})^2 + (5.4031\times 10^{-8})^2 (\Delta F^\text{ved}_{2s})^2 \\
			& + (-5.4031\times 10^{-8})^2 (\Delta F^\text{ved}_{3s})^2 \biggr]^{1/2},
		\end{aligned}
		\label{un_dr2}
	\end{equation}	

	\begin{equation}
		\begin{aligned}
			\Delta T_{M1} = \biggl[ &(4.8975)^2(\Delta R_{g,2s})^2 + (5.7316\times 10^{-5})^2(\Delta R_{m,2s})^2 + (-4.8606)^2(\Delta R_{g,3s})^2 \\
			& + (-3.4940\times 10^{-6})^2(\Delta R_{m,3s})^2 + (0.9630)^2(\Delta \epsilon_{g,1s})^2 + (1.1110\times 10^{-3})^2(\Delta \epsilon_{m,1s})^2 \\
			& + (-1.3522\times 10^{-3})^2(\Delta T^{(0)}_{2s})^2 + (5.1720\times 10^{-3})^2(\Delta T^{(0)}_{3s})^2 + (3.9110\times 10^{-3})^2(\Delta \omega_{\alpha 1})^2 \\
			& + (-1.3867\times 10^{-6})^2(\Delta \omega_{\alpha2})^2 + (2.2192\times 10^{-6})^2(\Delta \omega_{\alpha3})^2 + (-1.5076\times 10^{-2})^2(\Delta \omega_{\beta1})^2 \\
			& + (1.2558\times 10^{-6})^2(\Delta \omega_{\beta2})^2 + (-2.0092\times 10^{-6})^2(\Delta \omega_{\beta3})^2 + (-2.3306\times 10^{-8})^2 (\Delta F^\text{ved}_{2s})^2 \\
			& + (2.1101\times 10^{-8})^2 (\Delta F^\text{ved}_{3s})^2 \biggr]^{1/2}.
		\end{aligned}
		\label{un_tm1}
	\end{equation}	
\end{widetext}
{From these expressions, we obtain the explicit uncertainties for the five nuclear parameters, which are listed in the last column of Table I of the main text.}

{To test the sensitivity of our results to the assumed BW ratios for the $^{229}$Th$^{79+}$ ion, we vary both ratios $R_{g,3s}=1.095(5)$ and $R_{m,3s}=1.094(5)$ from 1.000(5) to 1.205(5). The resulting uncertainties for the five nuclear parameters are summarized in Table~\ref{uncertainties_R_scan2}. It is clearly seen that for $\omega_n$, the change in $\Delta \omega_n$ reaches 109 GHz; however, relative to its central value of 2 000 161 GHz, this change corresponds to only about 0.005\%, which is negligible for the present work. Similarly, the changes in the uncertainties of $\mu_g$, $\mu_m$ and $\delta\langle r^2\rangle_{gm}$ relative to their central values are all less than 0.6\%. The uncertainty in $T_{M1}$ changes by about 0.04 $\mu_N$, which is much smaller than our estimated uncertainty of 0.38 $\mu_N$. These tests indicate that the proposed extraction scheme is not significantly affected by possible differences between the Li-like and Na-like BW ratio factors.}

\begin{table*}[!htbp]
	\tabcolsep=0.0005cm
	\renewcommand\arraystretch{1.3}
	{\caption{
			Uncertainties of the nuclear parameters $\mu_g$, $\mu_m$, $\omega_n$, $\delta\langle r^2\rangle_{gm}$, and $T_{M1}$ as changes of $R_{g,3s}$ and $R_{m,3s}$ of the $^{229}$Th$^{79+}$ ion.}
		\label{uncertainties_R_scan2}}
	\begin{ruledtabular}
		\begin{tabular}{ccccccc}
			$R_{g,3s}$ & $R_{m,3s}$ & $\Delta\mu_g$ ($\mu_N$) & $\Delta\mu_m$ ($\mu_N$) & $\Delta\omega_n$ (GHz) & $\Delta\delta\langle r^2\rangle_{gm}$ (fm$^2$) & $\Delta T_{M1}$ ($\mu_N$) \\
			\hline
			1.005(5) & 1.000(5) & 0.0053 & 0.0070 & 1660 & 0.00081 & 0.41 \\
			1.025(5) & 1.020(5) & 0.0055 & 0.0072 & 1622 & 0.00079 & 0.40 \\
			1.045(5) & 1.040(5) & 0.0056 & 0.0074 & 1592 & 0.00078 & 0.39 \\
			1.065(5) & 1.060(5) & 0.0058 & 0.0076 & 1569 & 0.00077 & 0.38 \\
			1.085(5) & 1.080(5) & 0.0059 & 0.0078 & 1555 & 0.00076 & 0.38 \\
			1.095(5) & 1.094(5) & 0.0060 & 0.0080 & 1551 & 0.00076 & 0.38 \\
			1.125(5) & 1.128(5) & 0.0062 & 0.0083 & 1553 & 0.00076 & 0.38 \\
			1.145(5) & 1.140(5) & 0.0064 & 0.0085 & 1564 & 0.00077 & 0.38 \\
			1.165(5) & 1.160(5) & 0.0065 & 0.0087 & 1586 & 0.00078 & 0.39 \\
			1.185(5) & 1.180(5) & 0.0067 & 0.0089 & 1614 & 0.00079 & 0.40 \\
			1.205(5) & 1.200(5) & 0.0068 & 0.0091 & 1650 & 0.00081 & 0.42 \\
		\end{tabular}
	\end{ruledtabular}
\end{table*}

{Furthermore, to assess the impact of correlations not included in our previous estimation, we further consider possible correlations among the input uncertainties. Such correlations may arise from common physical effect, for example omitted higher-order QED terms in the electronic-structure quantities and the nuclear magnetization model entering the BW corrections. We therefore perform a covariance-aware uncertainty analysis. For any physical quantity \(P\) that depends on input quantities \(x_i\) and \(x_j\), the total uncertainty \(\Delta P\) is calculated by}
\begin{widetext}
	\begin{equation}
		\Delta P =
		\left[
		\sum_i
		\left( \frac{\partial P}{\partial x_i} \right)^2
		(\Delta x_i)^2
		+
		2\sum_{i<j}
		\rho_{ij}
		\left( \frac{\partial P}{\partial x_i} \right)
		\left( \frac{\partial P}{\partial x_j} \right)
		\Delta x_i \Delta x_j
		\right]^{1/2},
		\label{eq:covariance_propagation}
	\end{equation}
\end{widetext}
{
	where \(\rho_{ij}\) is a coefficient characterizing the degree of correlation between the uncertainties of \(x_i\) and \(x_j\). When \(\rho_{ij}=0\) for \(i\ne j\), it reduces to the uncorrelated analysis discussed above.
}

{
	In the correlated uncertainty analysis, we classify the input quantities into three categories according to their dominant uncertainty sources: (i) the QED-related quantities \(F_{2s}^{\mathrm{ved}}\), \(F_{3s}^{\mathrm{ved}}\), \(T_{2s}^{(0)}\), and \(T_{3s}^{(0)}\), which are mutually correlated; (ii) the BW-related quantities \(R_{x,2s}\), \(R_{x,3s}\), and \(\epsilon_{x,1s}\) (\(x=g,m\)), which are correlated within the same nuclear state; and (iii) the six transition frequencies, which are treated as correlated due to common systematic uncertainties. Correlations between different categories are neglected. Within each category, we take \(\rho_{ij}=1\). Starting from the case with no correlations, we successively include the correlations associated with QED corrections, BW corrections, and transition-frequency measurements. The resulting uncertainties are listed in Table~\ref{uncertainties_corr}.}

{
	The uncertainties of \(\mu_g\) and \(\mu_m\) are only weakly affected by these correlations. For example, \(\Delta\mu_g\) changes from \(0.00599\,\mu_N\) to \(0.00601\,\mu_N\), a change of less than 1\%. For \(\omega_n\), \(\delta\langle r^2\rangle_{gm}\), and \(T_{M1}\), the uncertainties actually decrease when correlations are properly accounted for. In particular, \(\Delta\omega_n\) decreases from \(1551\) GHz to \(192\) GHz when all three classes of correlations are included. Thus, the previous independent-error assumption yields a conservative upper bound for the uncertainties of \(\omega_n\), \(\delta\langle r^2\rangle_{gm}\), and \(T_{M1}\), and remains a good approximation for \(\mu_g\) and \(\mu_m\). }

\begin{table*}[!htbp]
	\tabcolsep=0.0005cm
	\renewcommand\arraystretch{1.3}
	{\caption{Uncertainties of the extracted nuclear parameters obtained by successively including correlations among different classes of input quantities. The units are \(\mu_N\) for \(\mu_g\), \(\mu_m\), and
			\(T_{M1}\), GHz for \(\omega_n\), and fm\(^2\) for
			\(\delta\langle r^2\rangle_{gm}\).}
		\label{uncertainties_corr}}
	\begin{ruledtabular}
		\begin{tabular}{lccccc}
			Correlations included & $\Delta \mu_g$ & $\Delta \mu_m$
			& $\Delta \omega_n$
			& $\Delta \delta\langle r^2\rangle_{gm}$
			& $\Delta T_{M1}$ \\
			\hline
			None (independent)                  & 0.00599 & 0.00800 & 1551 & 0.00080 & 0.38 \\
			QED                          & 0.00594 & 0.00794 & 357  & 0.00018 & 0.10 \\
			QED + BW                     & 0.00601 & 0.00805 & 315  & 0.00016 & 0.09 \\
			QED + BW + frequencies       & 0.00601 & 0.00805 & 192  & 0.00010 & 0.08 \\
		\end{tabular}
	\end{ruledtabular}
\end{table*}

\subsection{Experimental Feasibility of Highly Charged Thorium Ions}\label{S5}
The key experimental ingredients for hyperfine-structure (HFS) spectroscopy of highly charged thorium ions, namely the reliable production of the relevant charge states, long-lived confinement, and precision {{spectroscopic}} interrogation, have each been demonstrated in existing HCIs platforms.

Highly charged thorium ions in the range of interest have already been produced and spectroscopically characterized in electron-beam ion traps (EBITs)~\cite{beiersdorfer95,schneider91}. Precision x-ray spectroscopy of the thorium isonuclear sequence, for instance, has resolved the Li-like Th$^{87+}$ $2s_{1/2} \to 2p_{3/2}$ transition at $4025.23(14)$~eV, achieving a precision of 35~ppm for the line centers~\cite{beiersdorfer95}. These results establish both the availability of Th$^{87+}$ and the capability for high-resolution level identification in heavy thorium HCIs.

Long-lived confinement of heavy HCIs has also been demonstrated in complementary settings~\cite{beiersdorfer05,fritzsche05}. In cryogenic Penning traps, residual-gas pressures as low as $10^{-16}$~mbar yield storage times on the order of weeks, even for H-like U$^{91+}$~\cite{diederich99}. Storage rings offer an alternative: bare heavy ions can be injected, cooled, and decelerated for precision spectroscopy, as exemplified by a 4.6 eV accuracy determination of the ground-state Lamb shift in hydrogen-like uranium~\cite{gumberidze05}.

Most directly, laser HFS spectroscopy of heavy HCIs in a storage ring has been realized in the LIBELLE experiment at the GSI ESR~\cite{ullmann17,vollbrecht15}. By Doppler tuning the transition wavelength into the visible range, pulsed-laser scans across the predicted resonance enabled detection of the ground-state hyperfine splitting in H-like Bi$^{82+}$ and Li-like Bi$^{80+}$, providing reference data for both H-like and Li-like bismuth~\cite{ullmann17}. 
{This demonstrates a practical route for HFS measurements of heavy HCIs in storage rings.}

{
	For the $\beta1$ and $\beta2$ transitions of
	$^{229}\mathrm{Th}^{79+}$ shown in Fig.~1 of the main text, the
	corresponding splittings lie in the 7--8~THz range, where direct
	THz fluorescence detection from HCI beams or trapped HCIs is
	challenging. However, our scheme does not require such direct detection. The $\beta1$ splitting can be obtained from the VUV frequency
	difference: 	$\omega_{\beta1}=\omega_{\left|I_m,F=2\right\rangle\rightarrow\left|I_g,F=2\right\rangle}
	-\omega_{\left|I_m,F=2\right\rangle\rightarrow\left|I_g,F=3\right\rangle}$, and $\beta2$ splitting similarly.  
	Alternatively, THz-spaced splittings can be driven by femtosecond
	frequency-comb Raman spectroscopy, as demonstrated for a 1.8-THz
	Raman transition in a single trapped
	$^{40}\mathrm{Ca}^{+}$ ion~\cite{Solaro2018}. In trapped-ion
	implementations, the final-state readout can be performed by
	quantum-logic spectroscopy, as demonstrated for
	$\mathrm{Ar}^{13+}$~\cite{Micke2020}. Thus, although direct THz
	spectroscopy of trapped HCIs remains to be demonstrated, the
	essential ingredients for THz-range HFS measurements have been established.
}

Taken together, the demonstrated production and spectroscopic characterization of highly charged thorium ions in EBITs, the
established capability for long-term confinement of heavy HCIs, the
realization of laser HFS spectroscopy in storage rings, {and the
	availability of indirect THz-range spectroscopic approaches provide
	a practical basis for HFS measurements in the highly charged
	$^{229}$Th ions studied in this work. Ongoing efforts, such as the
	HiThor project at the ESR/HITRAP complex~\cite{Brandau2026},
	further support this perspective for precision spectroscopy of
	highly charged $^{229}$Th ions.}
\bibliography{positron}

\begin{thebibliography}{49}
\expandafter\ifx\csname natexlab\endcsname\relax\def\natexlab#1{#1}\fi
\expandafter\ifx\csname bibnamefont\endcsname\relax
  \def\bibnamefont#1{#1}\fi
\expandafter\ifx\csname bibfnamefont\endcsname\relax
  \def\bibfnamefont#1{#1}\fi
\expandafter\ifx\csname citenamefont\endcsname\relax
  \def\citenamefont#1{#1}\fi
\expandafter\ifx\csname url\endcsname\relax
  \def\url#1{\texttt{#1}}\fi
\expandafter\ifx\csname urlprefix\endcsname\relax\def\urlprefix{URL }\fi
\providecommand{\bibinfo}[2]{#2}
\providecommand{\eprint}[2][]{\url{#2}}

\bibitem[{\citenamefont{Peik and Tamm}(2003)}]{peik03a}
\bibinfo{author}{\bibfnamefont{E.}~\bibnamefont{Peik}} \bibnamefont{and}
  \bibinfo{author}{\bibfnamefont{C.}~\bibnamefont{Tamm}},
  \bibinfo{journal}{Europhys. Lett.} \textbf{\bibinfo{volume}{61}},
  \bibinfo{pages}{181} (\bibinfo{year}{2003}).

\bibitem[{\citenamefont{Zhang et~al.}(2024)\citenamefont{Zhang, Ooi, Higgins,
  Doyle, von~der Wense, Beeks, Leitner, Kazakov, Li, Thirol et~al.}}]{zhang24}
\bibinfo{author}{\bibfnamefont{C.~K.} \bibnamefont{Zhang}},
  \bibinfo{author}{\bibfnamefont{T.}~\bibnamefont{Ooi}},
  \bibinfo{author}{\bibfnamefont{J.~S.} \bibnamefont{Higgins}},
  \bibinfo{author}{\bibfnamefont{J.~F.} \bibnamefont{Doyle}},
  \bibinfo{author}{\bibfnamefont{L.}~\bibnamefont{von~der Wense}},
  \bibinfo{author}{\bibfnamefont{K.}~\bibnamefont{Beeks}},
  \bibinfo{author}{\bibfnamefont{A.}~\bibnamefont{Leitner}},
  \bibinfo{author}{\bibfnamefont{G.~A.} \bibnamefont{Kazakov}},
  \bibinfo{author}{\bibfnamefont{P.}~\bibnamefont{Li}},
  \bibinfo{author}{\bibfnamefont{P.~G.} \bibnamefont{Thirol}},
  \bibnamefont{et~al.}, \bibinfo{journal}{Nature}
  \textbf{\bibinfo{volume}{633}}, \bibinfo{pages}{63} (\bibinfo{year}{2024}).

\bibitem[{\citenamefont{Maheshwari and Jain}(2024)}]{maheshwari24}
\bibinfo{author}{\bibfnamefont{B.}~\bibnamefont{Maheshwari}} \bibnamefont{and}
  \bibinfo{author}{\bibfnamefont{A.~K.} \bibnamefont{Jain}},
  \bibinfo{journal}{Eur. Phys. J. Spec. Top.} \textbf{\bibinfo{volume}{233}},
  \bibinfo{pages}{1101} (\bibinfo{year}{2024}).

\bibitem[{\citenamefont{Von Der~Wense and Seiferle}(2020)}]{vonderwense20}
\bibinfo{author}{\bibfnamefont{L.}~\bibnamefont{Von Der~Wense}}
  \bibnamefont{and} \bibinfo{author}{\bibfnamefont{B.}~\bibnamefont{Seiferle}},
  \bibinfo{journal}{Eur. Phys. J. A} \textbf{\bibinfo{volume}{56}}
  (\bibinfo{year}{2020}).

\bibitem[{\citenamefont{Thirolf et~al.}(2024)\citenamefont{Thirolf, Kraemer,
  Moritz, and Scharl}}]{Thirolf24a}
\bibinfo{author}{\bibfnamefont{P.~G.} \bibnamefont{Thirolf}},
  \bibinfo{author}{\bibfnamefont{S.}~\bibnamefont{Kraemer}},
  \bibinfo{author}{\bibfnamefont{D.}~\bibnamefont{Moritz}}, \bibnamefont{and}
  \bibinfo{author}{\bibfnamefont{K.}~\bibnamefont{Scharl}},
  \bibinfo{journal}{Eur. Phys. J. Spec.Top.} \textbf{\bibinfo{volume}{233}},
  \bibinfo{pages}{1113} (\bibinfo{year}{2024}).

\bibitem[{\citenamefont{Peik et~al.}(2021)\citenamefont{Peik, Schumm,
  Safronova, Pálffy, Weitenberg, and Thirolf}}]{peik2021}
\bibinfo{author}{\bibfnamefont{E.}~\bibnamefont{Peik}},
  \bibinfo{author}{\bibfnamefont{T.}~\bibnamefont{Schumm}},
  \bibinfo{author}{\bibfnamefont{M.~S.} \bibnamefont{Safronova}},
  \bibinfo{author}{\bibfnamefont{A.}~\bibnamefont{Pálffy}},
  \bibinfo{author}{\bibfnamefont{J.}~\bibnamefont{Weitenberg}},
  \bibnamefont{and} \bibinfo{author}{\bibfnamefont{P.~G.}
  \bibnamefont{Thirolf}}, \bibinfo{journal}{Quantum Sci. Technol.}
  \textbf{\bibinfo{volume}{6}}, \bibinfo{pages}{034002} (\bibinfo{year}{2021}).

\bibitem[{\citenamefont{Caputo et~al.}(2025)\citenamefont{Caputo, Gazit,
  Hammer, Kopp, Paz, Perez, and Springmann}}]{caputo25}
\bibinfo{author}{\bibfnamefont{A.}~\bibnamefont{Caputo}},
  \bibinfo{author}{\bibfnamefont{D.}~\bibnamefont{Gazit}},
  \bibinfo{author}{\bibfnamefont{H.~W.} \bibnamefont{Hammer}},
  \bibinfo{author}{\bibfnamefont{J.}~\bibnamefont{Kopp}},
  \bibinfo{author}{\bibfnamefont{G.}~\bibnamefont{Paz}},
  \bibinfo{author}{\bibfnamefont{G.}~\bibnamefont{Perez}}, \bibnamefont{and}
  \bibinfo{author}{\bibfnamefont{K.}~\bibnamefont{Springmann}},
  \bibinfo{journal}{Phys. Rev. C} \textbf{\bibinfo{volume}{112}},
  \bibinfo{pages}{L031302} (\bibinfo{year}{2025}).

\bibitem[{\citenamefont{Fadeev et~al.}(2020)\citenamefont{Fadeev, Berengut, and
  Flambaum}}]{fadeev20}
\bibinfo{author}{\bibfnamefont{P.}~\bibnamefont{Fadeev}},
  \bibinfo{author}{\bibfnamefont{J.~C.} \bibnamefont{Berengut}},
  \bibnamefont{and} \bibinfo{author}{\bibfnamefont{V.~V.}
  \bibnamefont{Flambaum}}, \bibinfo{journal}{Phys. Rev. A}
  \textbf{\bibinfo{volume}{102}}, \bibinfo{pages}{052833}
  (\bibinfo{year}{2020}).

\bibitem[{\citenamefont{Zaheer et~al.}(2025)\citenamefont{Zaheer, Matjelo,
  Hume, Safronova, and Leibrandt}}]{zaheer25}
\bibinfo{author}{\bibfnamefont{M.~H.} \bibnamefont{Zaheer}},
  \bibinfo{author}{\bibfnamefont{N.~J.} \bibnamefont{Matjelo}},
  \bibinfo{author}{\bibfnamefont{D.~B.} \bibnamefont{Hume}},
  \bibinfo{author}{\bibfnamefont{M.~S.} \bibnamefont{Safronova}},
  \bibnamefont{and} \bibinfo{author}{\bibfnamefont{D.~R.}
  \bibnamefont{Leibrandt}}, \bibinfo{journal}{Phys. Rev. A}
  \textbf{\bibinfo{volume}{111}}, \bibinfo{pages}{012601}
  (\bibinfo{year}{2025}).

\bibitem[{\citenamefont{Perera et~al.}(2025)\citenamefont{Perera, Morgan,
  Hudson, and Derevianko}}]{perera25a}
\bibinfo{author}{\bibfnamefont{U.~C.} \bibnamefont{Perera}},
  \bibinfo{author}{\bibfnamefont{H.~W.~T.} \bibnamefont{Morgan}},
  \bibinfo{author}{\bibfnamefont{E.~R.} \bibnamefont{Hudson}},
  \bibnamefont{and}
  \bibinfo{author}{\bibfnamefont{A.}~\bibnamefont{Derevianko}},
  \bibinfo{journal}{Phys. Rev. Lett.} \textbf{\bibinfo{volume}{135}},
  \bibinfo{pages}{123001} (\bibinfo{year}{2025}).

\bibitem[{\citenamefont{Safronova et~al.}(2018)\citenamefont{Safronova, Porsev,
  Kozlov, Thielking, Okhapkin, G\l{}owacki, Meier, and Peik}}]{safronova18a}
\bibinfo{author}{\bibfnamefont{M.~S.} \bibnamefont{Safronova}},
  \bibinfo{author}{\bibfnamefont{S.~G.} \bibnamefont{Porsev}},
  \bibinfo{author}{\bibfnamefont{M.~G.} \bibnamefont{Kozlov}},
  \bibinfo{author}{\bibfnamefont{J.}~\bibnamefont{Thielking}},
  \bibinfo{author}{\bibfnamefont{M.~V.} \bibnamefont{Okhapkin}},
  \bibinfo{author}{\bibfnamefont{P.}~\bibnamefont{G\l{}owacki}},
  \bibinfo{author}{\bibfnamefont{D.~M.} \bibnamefont{Meier}}, \bibnamefont{and}
  \bibinfo{author}{\bibfnamefont{E.}~\bibnamefont{Peik}},
  \bibinfo{journal}{Phys. Rev. Lett.} \textbf{\bibinfo{volume}{121}},
  \bibinfo{pages}{213001} (\bibinfo{year}{2018}).

\bibitem[{\citenamefont{Porsev and Flambaum}(2010{\natexlab{a}})}]{porsev10a}
\bibinfo{author}{\bibfnamefont{S.~G.} \bibnamefont{Porsev}} \bibnamefont{and}
  \bibinfo{author}{\bibfnamefont{V.~V.} \bibnamefont{Flambaum}},
  \bibinfo{journal}{Phys. Rev. A} \textbf{\bibinfo{volume}{81}},
  \bibinfo{pages}{032504} (\bibinfo{year}{2010}{\natexlab{a}}).

\bibitem[{\citenamefont{Minkov and P\'alffy}(2017)}]{minkov17}
\bibinfo{author}{\bibfnamefont{N.}~\bibnamefont{Minkov}} \bibnamefont{and}
  \bibinfo{author}{\bibfnamefont{A.}~\bibnamefont{P\'alffy}},
  \bibinfo{journal}{Phys. Rev. Lett.} \textbf{\bibinfo{volume}{118}},
  \bibinfo{pages}{212501} (\bibinfo{year}{2017}).

\bibitem[{\citenamefont{Bilous et~al.}(2018)\citenamefont{Bilous, Minkov, and
  P\'alffy}}]{bilous18b}
\bibinfo{author}{\bibfnamefont{P.~V.} \bibnamefont{Bilous}},
  \bibinfo{author}{\bibfnamefont{N.}~\bibnamefont{Minkov}}, \bibnamefont{and}
  \bibinfo{author}{\bibfnamefont{A.}~\bibnamefont{P\'alffy}},
  \bibinfo{journal}{Phys. Rev. C} \textbf{\bibinfo{volume}{97}},
  \bibinfo{pages}{044320} (\bibinfo{year}{2018}).

\bibitem[{\citenamefont{Tiedau et~al.}(2024)\citenamefont{Tiedau, Okhapkin,
  Zhang, Thielking, Zitzer, Peik, Schaden, Pronebner, Morawetz, De~Col
  et~al.}}]{tiedau24}
\bibinfo{author}{\bibfnamefont{J.}~\bibnamefont{Tiedau}},
  \bibinfo{author}{\bibfnamefont{M.~V.} \bibnamefont{Okhapkin}},
  \bibinfo{author}{\bibfnamefont{K.}~\bibnamefont{Zhang}},
  \bibinfo{author}{\bibfnamefont{J.}~\bibnamefont{Thielking}},
  \bibinfo{author}{\bibfnamefont{G.}~\bibnamefont{Zitzer}},
  \bibinfo{author}{\bibfnamefont{E.}~\bibnamefont{Peik}},
  \bibinfo{author}{\bibfnamefont{F.}~\bibnamefont{Schaden}},
  \bibinfo{author}{\bibfnamefont{T.}~\bibnamefont{Pronebner}},
  \bibinfo{author}{\bibfnamefont{I.}~\bibnamefont{Morawetz}},
  \bibinfo{author}{\bibfnamefont{L.~T.} \bibnamefont{De~Col}},
  \bibnamefont{et~al.}, \bibinfo{journal}{Phys. Rev. Lett.}
  \textbf{\bibinfo{volume}{132}}, \bibinfo{pages}{182501}
  (\bibinfo{year}{2024}).

\bibitem[{\citenamefont{Shabaev et~al.}(2022)\citenamefont{Shabaev, Glazov,
  Ryzhkov, Brandau, Plunien, Quint, Volchkova, and Zinenko}}]{shabaev22}
\bibinfo{author}{\bibfnamefont{V.~M.} \bibnamefont{Shabaev}},
  \bibinfo{author}{\bibfnamefont{D.~A.} \bibnamefont{Glazov}},
  \bibinfo{author}{\bibfnamefont{A.~M.} \bibnamefont{Ryzhkov}},
  \bibinfo{author}{\bibfnamefont{C.}~\bibnamefont{Brandau}},
  \bibinfo{author}{\bibfnamefont{G.}~\bibnamefont{Plunien}},
  \bibinfo{author}{\bibfnamefont{W.}~\bibnamefont{Quint}},
  \bibinfo{author}{\bibfnamefont{A.~M.} \bibnamefont{Volchkova}},
  \bibnamefont{and} \bibinfo{author}{\bibfnamefont{D.~V.}
  \bibnamefont{Zinenko}}, \bibinfo{journal}{Phys. Rev. Lett.}
  \textbf{\bibinfo{volume}{128}}, \bibinfo{pages}{043001}
  (\bibinfo{year}{2022}).

\bibitem[{\citenamefont{Kozlov et~al.}(2024)\citenamefont{Kozlov, Oleynichenko,
  Budker, Glazov, Lomachuk, Shabaev, Titov, Tupitsyn, and Volotka}}]{kozlov24}
\bibinfo{author}{\bibfnamefont{M.~G.} \bibnamefont{Kozlov}},
  \bibinfo{author}{\bibfnamefont{A.~V.} \bibnamefont{Oleynichenko}},
  \bibinfo{author}{\bibfnamefont{D.}~\bibnamefont{Budker}},
  \bibinfo{author}{\bibfnamefont{D.~A.} \bibnamefont{Glazov}},
  \bibinfo{author}{\bibfnamefont{Y.~V.} \bibnamefont{Lomachuk}},
  \bibinfo{author}{\bibfnamefont{V.~M.} \bibnamefont{Shabaev}},
  \bibinfo{author}{\bibfnamefont{A.~V.} \bibnamefont{Titov}},
  \bibinfo{author}{\bibfnamefont{I.~I.} \bibnamefont{Tupitsyn}},
  \bibnamefont{and} \bibinfo{author}{\bibfnamefont{A.~V.}
  \bibnamefont{Volotka}}, \bibinfo{journal}{Phys. Rev. A}
  \textbf{\bibinfo{volume}{109}}, \bibinfo{pages}{042806}
  (\bibinfo{year}{2024}).

\bibitem[{\citenamefont{Beloy}(2014)}]{beloy14}
\bibinfo{author}{\bibfnamefont{K.}~\bibnamefont{Beloy}},
  \bibinfo{journal}{Phys. Rev. Lett.} \textbf{\bibinfo{volume}{112}},
  \bibinfo{pages}{062503} (\bibinfo{year}{2014}).

\bibitem[{\citenamefont{Porsev and Flambaum}(2010{\natexlab{b}})}]{porsev10}
\bibinfo{author}{\bibfnamefont{S.~G.} \bibnamefont{Porsev}} \bibnamefont{and}
  \bibinfo{author}{\bibfnamefont{V.~V.} \bibnamefont{Flambaum}},
  \bibinfo{journal}{Phys. Rev. A} \textbf{\bibinfo{volume}{81}},
  \bibinfo{pages}{032504} (\bibinfo{year}{2010}{\natexlab{b}}).

\bibitem[{\citenamefont{Bohr and Weisskopf}(1950)}]{bohr50}
\bibinfo{author}{\bibfnamefont{A.}~\bibnamefont{Bohr}} \bibnamefont{and}
  \bibinfo{author}{\bibfnamefont{V.~F.} \bibnamefont{Weisskopf}},
  \bibinfo{journal}{Phys. Rev.} \textbf{\bibinfo{volume}{77}},
  \bibinfo{pages}{94} (\bibinfo{year}{1950}).

\bibitem[{\citenamefont{Shabaev}(1999)}]{shabaev99hfs}
\bibinfo{author}{\bibfnamefont{V.~M.} \bibnamefont{Shabaev}}, in
  \emph{\bibinfo{booktitle}{Atomic Physics with Heavy Ions}}, edited by
  \bibinfo{editor}{\bibfnamefont{H.~F.} \bibnamefont{Beyer}} \bibnamefont{and}
  \bibinfo{editor}{\bibfnamefont{V.~P.} \bibnamefont{Shevelko}}
  (\bibinfo{publisher}{Springer}, \bibinfo{address}{Berlin},
  \bibinfo{year}{1999}), vol.~\bibinfo{volume}{26} of
  \emph{\bibinfo{series}{Springer Series on Atoms and Plasmas}}, pp.
  \bibinfo{pages}{139--159}.

\bibitem[{\citenamefont{Tkalya and Nikolaev}(2016)}]{tkalya16}
\bibinfo{author}{\bibfnamefont{E.~V.} \bibnamefont{Tkalya}} \bibnamefont{and}
  \bibinfo{author}{\bibfnamefont{A.~V.} \bibnamefont{Nikolaev}},
  \bibinfo{journal}{Phys. Rev. C} \textbf{\bibinfo{volume}{94}},
  \bibinfo{pages}{014323} (\bibinfo{year}{2016}).

\bibitem[{\citenamefont{Ullmann et~al.}(2017)\citenamefont{Ullmann, Andelkovic,
  Brandau, Dax, Geithner, Geppert, Gorges, Hammen, Hannen, Kaufmann
  et~al.}}]{ullmann17}
\bibinfo{author}{\bibfnamefont{J.}~\bibnamefont{Ullmann}},
  \bibinfo{author}{\bibfnamefont{Z.}~\bibnamefont{Andelkovic}},
  \bibinfo{author}{\bibfnamefont{C.}~\bibnamefont{Brandau}},
  \bibinfo{author}{\bibfnamefont{A.}~\bibnamefont{Dax}},
  \bibinfo{author}{\bibfnamefont{W.}~\bibnamefont{Geithner}},
  \bibinfo{author}{\bibfnamefont{C.}~\bibnamefont{Geppert}},
  \bibinfo{author}{\bibfnamefont{C.}~\bibnamefont{Gorges}},
  \bibinfo{author}{\bibfnamefont{M.}~\bibnamefont{Hammen}},
  \bibinfo{author}{\bibfnamefont{V.}~\bibnamefont{Hannen}},
  \bibinfo{author}{\bibfnamefont{S.}~\bibnamefont{Kaufmann}},
  \bibnamefont{et~al.}, \bibinfo{journal}{Nat. Commun.}
  \textbf{\bibinfo{volume}{8}}, \bibinfo{pages}{15484} (\bibinfo{year}{2017}).

\bibitem[{\citenamefont{Shabaev et~al.}(2001)\citenamefont{Shabaev, Artemyev,
  Yerokhin, Zherebtsov, and Soff}}]{shabaev01}
\bibinfo{author}{\bibfnamefont{V.~M.} \bibnamefont{Shabaev}},
  \bibinfo{author}{\bibfnamefont{A.~N.} \bibnamefont{Artemyev}},
  \bibinfo{author}{\bibfnamefont{V.~A.} \bibnamefont{Yerokhin}},
  \bibinfo{author}{\bibfnamefont{O.~M.} \bibnamefont{Zherebtsov}},
  \bibnamefont{and} \bibinfo{author}{\bibfnamefont{G.}~\bibnamefont{Soff}},
  \bibinfo{journal}{Phys. Rev. Lett.} \textbf{\bibinfo{volume}{86}},
  \bibinfo{pages}{3959} (\bibinfo{year}{2001}).

\bibitem[{\citenamefont{Roberts et~al.}(2022)\citenamefont{Roberts, Ranclaud,
  and Ginges}}]{roberts22}
\bibinfo{author}{\bibfnamefont{B.~M.} \bibnamefont{Roberts}},
  \bibinfo{author}{\bibfnamefont{P.~G.} \bibnamefont{Ranclaud}},
  \bibnamefont{and} \bibinfo{author}{\bibfnamefont{J.~S.~M.}
  \bibnamefont{Ginges}}, \bibinfo{journal}{Phys. Rev. A}
  \textbf{\bibinfo{volume}{105}}, \bibinfo{pages}{052802}
  (\bibinfo{year}{2022}).

\bibitem[{\citenamefont{Ekman et~al.}(2019)\citenamefont{Ekman, Jönsson,
  Godefroid, Nazé, Gaigalas, and Bieroń}}]{ekman19a}
\bibinfo{author}{\bibfnamefont{J.}~\bibnamefont{Ekman}},
  \bibinfo{author}{\bibfnamefont{P.}~\bibnamefont{Jönsson}},
  \bibinfo{author}{\bibfnamefont{M.}~\bibnamefont{Godefroid}},
  \bibinfo{author}{\bibfnamefont{C.}~\bibnamefont{Nazé}},
  \bibinfo{author}{\bibfnamefont{G.}~\bibnamefont{Gaigalas}}, \bibnamefont{and}
  \bibinfo{author}{\bibfnamefont{J.}~\bibnamefont{Bieroń}},
  \bibinfo{journal}{Comput. Phys. Commun.} \textbf{\bibinfo{volume}{235}},
  \bibinfo{pages}{433} (\bibinfo{year}{2019}).

\bibitem[{\citenamefont{{Froese Fischer} et~al.}(2019)\citenamefont{{Froese
  Fischer}, Gaigalas, Jönsson, and Bieroń}}]{Fischer19a}
\bibinfo{author}{\bibfnamefont{C.}~\bibnamefont{{Froese Fischer}}},
  \bibinfo{author}{\bibfnamefont{G.}~\bibnamefont{Gaigalas}},
  \bibinfo{author}{\bibfnamefont{P.}~\bibnamefont{Jönsson}}, \bibnamefont{and}
  \bibinfo{author}{\bibfnamefont{J.}~\bibnamefont{Bieroń}},
  \bibinfo{journal}{Comput. Phys. Commun.} \textbf{\bibinfo{volume}{237}},
  \bibinfo{pages}{184} (\bibinfo{year}{2019}).

\bibitem[{\citenamefont{Porsev et~al.}(2021)\citenamefont{Porsev, Safronova,
  and Kozlov}}]{Porsev21a}
\bibinfo{author}{\bibfnamefont{S.~G.} \bibnamefont{Porsev}},
  \bibinfo{author}{\bibfnamefont{M.~S.} \bibnamefont{Safronova}},
  \bibnamefont{and} \bibinfo{author}{\bibfnamefont{M.~G.}
  \bibnamefont{Kozlov}}, \bibinfo{journal}{Phys. Rev. Lett.}
  \textbf{\bibinfo{volume}{127}}, \bibinfo{pages}{253001}
  (\bibinfo{year}{2021}).

\bibitem[{\citenamefont{Yamaguchi et~al.}(2024)\citenamefont{Yamaguchi,
  Shigekawa, Haba, Kikunaga, Shirasaki, Wada, and Katori}}]{katori24a}
\bibinfo{author}{\bibfnamefont{A.}~\bibnamefont{Yamaguchi}},
  \bibinfo{author}{\bibfnamefont{Y.}~\bibnamefont{Shigekawa}},
  \bibinfo{author}{\bibfnamefont{H.}~\bibnamefont{Haba}},
  \bibinfo{author}{\bibfnamefont{H.}~\bibnamefont{Kikunaga}},
  \bibinfo{author}{\bibfnamefont{K.}~\bibnamefont{Shirasaki}},
  \bibinfo{author}{\bibfnamefont{M.}~\bibnamefont{Wada}}, \bibnamefont{and}
  \bibinfo{author}{\bibfnamefont{H.}~\bibnamefont{Katori}},
  \bibinfo{journal}{Nature} \textbf{\bibinfo{volume}{629}}, \bibinfo{pages}{62}
  (\bibinfo{year}{2024}).

\bibitem[{\citenamefont{Morgan et~al.}(2025)\citenamefont{Morgan, Tan, Elwell,
  Alexandrova, Hudson, and Derevianko}}]{morgan25}
\bibinfo{author}{\bibfnamefont{H.~W.~T.} \bibnamefont{Morgan}},
  \bibinfo{author}{\bibfnamefont{H.~B.~T.} \bibnamefont{Tan}},
  \bibinfo{author}{\bibfnamefont{R.}~\bibnamefont{Elwell}},
  \bibinfo{author}{\bibfnamefont{A.~N.} \bibnamefont{Alexandrova}},
  \bibinfo{author}{\bibfnamefont{E.~R.} \bibnamefont{Hudson}},
  \bibnamefont{and}
  \bibinfo{author}{\bibfnamefont{A.}~\bibnamefont{Derevianko}},
  \bibinfo{journal}{Phys. Rev. Lett.} \textbf{\bibinfo{volume}{134}},
  \bibinfo{pages}{253801} (\bibinfo{year}{2025}).

\bibitem[{\citenamefont{Schwartz}(1955)}]{Schwartz55a}
\bibinfo{author}{\bibfnamefont{C.}~\bibnamefont{Schwartz}},
  \bibinfo{journal}{Phys. Rev.} \textbf{\bibinfo{volume}{97}},
  \bibinfo{pages}{380} (\bibinfo{year}{1955}).

\bibitem[{\citenamefont{Wang and Wang}(2023)}]{wang23}
\bibinfo{author}{\bibfnamefont{W.}~\bibnamefont{Wang}} \bibnamefont{and}
  \bibinfo{author}{\bibfnamefont{X.}~\bibnamefont{Wang}},
  \bibinfo{journal}{Phys. Rev. Res.} \textbf{\bibinfo{volume}{5}},
  \bibinfo{pages}{043232} (\bibinfo{year}{2023}).

\bibitem[{\citenamefont{Edmonds}(1957)}]{edmonds1957}
\bibinfo{author}{\bibfnamefont{A.~R.} \bibnamefont{Edmonds}},
  \emph{\bibinfo{title}{Angular Momentum in Quantum Mechanics}}
  (\bibinfo{publisher}{Princeton University Press},
  \bibinfo{address}{Princeton, NJ}, \bibinfo{year}{1957}).

\bibitem[{\citenamefont{Trees}(1953)}]{trees58}
\bibinfo{author}{\bibfnamefont{R.~E.} \bibnamefont{Trees}},
  \bibinfo{journal}{Phys. Rev.} \textbf{\bibinfo{volume}{92}},
  \bibinfo{pages}{308} (\bibinfo{year}{1953}).

\bibitem[{\citenamefont{Blaise and Wyart}(2021)}]{th2024}
\bibinfo{author}{\bibfnamefont{A.}~\bibnamefont{Blaise}} \bibnamefont{and}
  \bibinfo{author}{\bibfnamefont{J.}~\bibnamefont{Wyart}}
  (\bibinfo{year}{2021}),
  \bibinfo{note}{\url{http://www.lac.universite-paris-saclay.fr/Data/Database/}}.

\bibitem[{\citenamefont{Torbohm et~al.}(1985)\citenamefont{Torbohm, Fricke, and
  Ros\'en}}]{torbohm85a}
\bibinfo{author}{\bibfnamefont{G.}~\bibnamefont{Torbohm}},
  \bibinfo{author}{\bibfnamefont{B.}~\bibnamefont{Fricke}}, \bibnamefont{and}
  \bibinfo{author}{\bibfnamefont{A.}~\bibnamefont{Ros\'en}},
  \bibinfo{journal}{Phys. Rev. A} \textbf{\bibinfo{volume}{31}},
  \bibinfo{pages}{2038} (\bibinfo{year}{1985}).

\bibitem[{\citenamefont{Zheng et~al.}(2026)\citenamefont{Zheng, Xu, Li, Zhang,
  Shi, and Tang}}]{zheng26}
\bibinfo{author}{\bibfnamefont{H.~Y.} \bibnamefont{Zheng}},
  \bibinfo{author}{\bibfnamefont{Y.~L.} \bibnamefont{Xu}},
  \bibinfo{author}{\bibfnamefont{J.~G.} \bibnamefont{Li}},
  \bibinfo{author}{\bibfnamefont{Y.~H.} \bibnamefont{Zhang}},
  \bibinfo{author}{\bibfnamefont{T.~Y.} \bibnamefont{Shi}}, \bibnamefont{and}
  \bibinfo{author}{\bibfnamefont{L.~Y.} \bibnamefont{Tang}},
  \bibinfo{journal}{Chin. Phys. Lett.} \textbf{\bibinfo{volume}{43}},
  \bibinfo{pages}{020301} (\bibinfo{year}{2026}).

\bibitem[{\citenamefont{Parpia and Mohanty}(1992)}]{parpia92a}
\bibinfo{author}{\bibfnamefont{F.~A.} \bibnamefont{Parpia}} \bibnamefont{and}
  \bibinfo{author}{\bibfnamefont{A.~K.} \bibnamefont{Mohanty}},
  \bibinfo{journal}{Phys. Rev. A} \textbf{\bibinfo{volume}{46}},
  \bibinfo{pages}{3735} (\bibinfo{year}{1992}).

\bibitem[{\citenamefont{Ginges and Volotka}(2018)}]{ginges18}
\bibinfo{author}{\bibfnamefont{J.~S.~M.} \bibnamefont{Ginges}}
  \bibnamefont{and} \bibinfo{author}{\bibfnamefont{A.~V.}
  \bibnamefont{Volotka}}, \bibinfo{journal}{Phys. Rev. A}
  \textbf{\bibinfo{volume}{98}}, \bibinfo{pages}{032504}
  (\bibinfo{year}{2018}).

\bibitem[{\citenamefont{Beiersdorfer et~al.}(1995)\citenamefont{Beiersdorfer,
  Osterheld, Elliott, Chen, Knapp, and Reed}}]{beiersdorfer95}
\bibinfo{author}{\bibfnamefont{P.}~\bibnamefont{Beiersdorfer}},
  \bibinfo{author}{\bibfnamefont{A.}~\bibnamefont{Osterheld}},
  \bibinfo{author}{\bibfnamefont{S.~R.} \bibnamefont{Elliott}},
  \bibinfo{author}{\bibfnamefont{M.~H.} \bibnamefont{Chen}},
  \bibinfo{author}{\bibfnamefont{D.}~\bibnamefont{Knapp}}, \bibnamefont{and}
  \bibinfo{author}{\bibfnamefont{K.}~\bibnamefont{Reed}},
  \bibinfo{journal}{Phys. Rev. A} \textbf{\bibinfo{volume}{52}},
  \bibinfo{pages}{2693} (\bibinfo{year}{1995}).

\bibitem[{\citenamefont{Schneider et~al.}(1991)\citenamefont{Schneider, Clark,
  Penetrante, McDonald, DeWitt, and Bardsley}}]{schneider91}
\bibinfo{author}{\bibfnamefont{D.}~\bibnamefont{Schneider}},
  \bibinfo{author}{\bibfnamefont{M.~W.} \bibnamefont{Clark}},
  \bibinfo{author}{\bibfnamefont{B.~M.} \bibnamefont{Penetrante}},
  \bibinfo{author}{\bibfnamefont{J.}~\bibnamefont{McDonald}},
  \bibinfo{author}{\bibfnamefont{D.}~\bibnamefont{DeWitt}}, \bibnamefont{and}
  \bibinfo{author}{\bibfnamefont{J.~N.} \bibnamefont{Bardsley}},
  \bibinfo{journal}{Phys. Rev. A} \textbf{\bibinfo{volume}{44}},
  \bibinfo{pages}{3119} (\bibinfo{year}{1991}).

\bibitem[{\citenamefont{Beiersdorfer et~al.}(2005)\citenamefont{Beiersdorfer,
  Chen, Thorn, and Tr\"abert}}]{beiersdorfer05}
\bibinfo{author}{\bibfnamefont{P.}~\bibnamefont{Beiersdorfer}},
  \bibinfo{author}{\bibfnamefont{H.}~\bibnamefont{Chen}},
  \bibinfo{author}{\bibfnamefont{D.~B.} \bibnamefont{Thorn}}, \bibnamefont{and}
  \bibinfo{author}{\bibfnamefont{E.}~\bibnamefont{Tr\"abert}},
  \bibinfo{journal}{Phys. Rev. Lett.} \textbf{\bibinfo{volume}{95}},
  \bibinfo{pages}{233003} (\bibinfo{year}{2005}).

\bibitem[{\citenamefont{Fritzsche et~al.}(2005)\citenamefont{Fritzsche,
  Indelicato, and Stöhlker}}]{fritzsche05}
\bibinfo{author}{\bibfnamefont{S.}~\bibnamefont{Fritzsche}},
  \bibinfo{author}{\bibfnamefont{P.}~\bibnamefont{Indelicato}},
  \bibnamefont{and}
  \bibinfo{author}{\bibfnamefont{T.}~\bibnamefont{Stöhlker}},
  \bibinfo{journal}{J. Phys. B: At. Mol. Opt. Phys.}
  \textbf{\bibinfo{volume}{38}}, \bibinfo{pages}{S707} (\bibinfo{year}{2005}).

\bibitem[{\citenamefont{Diederich et~al.}(1999)\citenamefont{Diederich,
  H{\"a}ffner, Hermanspahn, Immel, Kluge, Ley, Mann, Stahl, Quint, Verd{\'u}
  et~al.}}]{diederich99}
\bibinfo{author}{\bibfnamefont{M.}~\bibnamefont{Diederich}},
  \bibinfo{author}{\bibfnamefont{H.}~\bibnamefont{H{\"a}ffner}},
  \bibinfo{author}{\bibfnamefont{N.}~\bibnamefont{Hermanspahn}},
  \bibinfo{author}{\bibfnamefont{M.}~\bibnamefont{Immel}},
  \bibinfo{author}{\bibfnamefont{H.~J.} \bibnamefont{Kluge}},
  \bibinfo{author}{\bibfnamefont{R.}~\bibnamefont{Ley}},
  \bibinfo{author}{\bibfnamefont{R.}~\bibnamefont{Mann}},
  \bibinfo{author}{\bibfnamefont{S.}~\bibnamefont{Stahl}},
  \bibinfo{author}{\bibfnamefont{W.}~\bibnamefont{Quint}},
  \bibinfo{author}{\bibfnamefont{J.}~\bibnamefont{Verd{\'u}}},
  \bibnamefont{et~al.}, in \emph{\bibinfo{booktitle}{Trapped Charged Particles
  and Fundamental Physics}} (\bibinfo{address}{Asilomar, California (USA)},
  \bibinfo{year}{1999}), pp. \bibinfo{pages}{43--51}.

\bibitem[{\citenamefont{Gumberidze et~al.}(2005)\citenamefont{Gumberidze,
  St\"ohlker, Bana\ifmmode~\acute{s}\else \'{s}\fi{}, Beckert, Beller, Beyer,
  Bosch, Hagmann, Kozhuharov, Liesen et~al.}}]{gumberidze05}
\bibinfo{author}{\bibfnamefont{A.}~\bibnamefont{Gumberidze}},
  \bibinfo{author}{\bibfnamefont{T.}~\bibnamefont{St\"ohlker}},
  \bibinfo{author}{\bibfnamefont{D.}~\bibnamefont{Bana\ifmmode~\acute{s}\else
  \'{s}\fi{}}}, \bibinfo{author}{\bibfnamefont{K.}~\bibnamefont{Beckert}},
  \bibinfo{author}{\bibfnamefont{P.}~\bibnamefont{Beller}},
  \bibinfo{author}{\bibfnamefont{H.~F.} \bibnamefont{Beyer}},
  \bibinfo{author}{\bibfnamefont{F.}~\bibnamefont{Bosch}},
  \bibinfo{author}{\bibfnamefont{S.}~\bibnamefont{Hagmann}},
  \bibinfo{author}{\bibfnamefont{C.}~\bibnamefont{Kozhuharov}},
  \bibinfo{author}{\bibfnamefont{D.}~\bibnamefont{Liesen}},
  \bibnamefont{et~al.}, \bibinfo{journal}{Phys. Rev. Lett.}
  \textbf{\bibinfo{volume}{94}}, \bibinfo{pages}{223001}
  (\bibinfo{year}{2005}).

\bibitem[{\citenamefont{Vollbrecht et~al.}(2015)\citenamefont{Vollbrecht,
  Andelkovic, Dax, Geithner, Geppert, Gorges, Hammen, Hannen, Kaufmann, König
  et~al.}}]{vollbrecht15}
\bibinfo{author}{\bibfnamefont{J.}~\bibnamefont{Vollbrecht}},
  \bibinfo{author}{\bibfnamefont{Z.}~\bibnamefont{Andelkovic}},
  \bibinfo{author}{\bibfnamefont{A.}~\bibnamefont{Dax}},
  \bibinfo{author}{\bibfnamefont{W.}~\bibnamefont{Geithner}},
  \bibinfo{author}{\bibfnamefont{C.}~\bibnamefont{Geppert}},
  \bibinfo{author}{\bibfnamefont{C.}~\bibnamefont{Gorges}},
  \bibinfo{author}{\bibfnamefont{M.}~\bibnamefont{Hammen}},
  \bibinfo{author}{\bibfnamefont{V.}~\bibnamefont{Hannen}},
  \bibinfo{author}{\bibfnamefont{S.}~\bibnamefont{Kaufmann}},
  \bibinfo{author}{\bibfnamefont{K.}~\bibnamefont{König}},
  \bibnamefont{et~al.}, \bibinfo{journal}{J. Phys.: Conf. Ser.}
  \textbf{\bibinfo{volume}{583}}, \bibinfo{pages}{012002}
  (\bibinfo{year}{2015}).

\bibitem[{\citenamefont{Solaro et~al.}(2018)\citenamefont{Solaro, Meyer,
  Fisher, DePalatis, and Drewsen}}]{Solaro2018}
\bibinfo{author}{\bibfnamefont{C.}~\bibnamefont{Solaro}},
  \bibinfo{author}{\bibfnamefont{S.}~\bibnamefont{Meyer}},
  \bibinfo{author}{\bibfnamefont{K.}~\bibnamefont{Fisher}},
  \bibinfo{author}{\bibfnamefont{M.~V.} \bibnamefont{DePalatis}},
  \bibnamefont{and} \bibinfo{author}{\bibfnamefont{M.}~\bibnamefont{Drewsen}},
  \bibinfo{journal}{Phys. Rev. Lett.} \textbf{\bibinfo{volume}{120}},
  \bibinfo{pages}{253601} (\bibinfo{year}{2018}).

\bibitem[{\citenamefont{Micke et~al.}(2020)\citenamefont{Micke, Leopold, King,
  Benkler, Spie{\ss}, Schm{\"o}ger, Schwarz, Crespo L{\'o}pez-Urrutia, and
  Schmidt}}]{Micke2020}
\bibinfo{author}{\bibfnamefont{P.}~\bibnamefont{Micke}},
  \bibinfo{author}{\bibfnamefont{T.}~\bibnamefont{Leopold}},
  \bibinfo{author}{\bibfnamefont{S.~A.} \bibnamefont{King}},
  \bibinfo{author}{\bibfnamefont{E.}~\bibnamefont{Benkler}},
  \bibinfo{author}{\bibfnamefont{L.~J.} \bibnamefont{Spie{\ss}}},
  \bibinfo{author}{\bibfnamefont{L.}~\bibnamefont{Schm{\"o}ger}},
  \bibinfo{author}{\bibfnamefont{M.}~\bibnamefont{Schwarz}},
  \bibinfo{author}{\bibfnamefont{J.~R.} \bibnamefont{Crespo
  L{\'o}pez-Urrutia}}, \bibnamefont{and} \bibinfo{author}{\bibfnamefont{P.~O.}
  \bibnamefont{Schmidt}}, \bibinfo{journal}{Nature}
  \textbf{\bibinfo{volume}{578}}, \bibinfo{pages}{60} (\bibinfo{year}{2020}).

\bibitem[{\citenamefont{Brandau and St{\"o}hlker}(2026)}]{Brandau2026}
\bibinfo{author}{\bibfnamefont{C.}~\bibnamefont{Brandau}} \bibnamefont{and}
  \bibinfo{author}{\bibfnamefont{T.}~\bibnamefont{St{\"o}hlker}}, in
  \emph{\bibinfo{booktitle}{Book of Abstracts: International Conference on
  Precision Physics of Simple Atomic Systems}} (\bibinfo{address}{Vienna,
  Austria}, \bibinfo{year}{2026}), p.~\bibinfo{pages}{43},
  \bibinfo{note}{18--22 May 2026, {\"O}AW}.

\end{thebibliography}

\end{document}